\documentclass[a4paper,11pt]{article}

\usepackage{jinstpub} 
\usepackage{bm}
\usepackage{multirow} 
\usepackage{url}
\usepackage{color}
\usepackage[normalem]{ulem}
\usepackage[T1]{fontenc}
\usepackage[dvipsnames,table]{xcolor}
\usepackage{ragged2e}

\hypersetup{urlcolor=RedViolet,
	    citecolor=ForestGreen ,
	    linkcolor=RoyalBlue, 
	    colorlinks=true
}

\title{Directional detection of keV proton and carbon recoils with MIMAC}

\author[a]{C. Beaufort,}
\author[a]{O. Guillaudin,}
\author[a]{D. Santos,}
\author[a]{N. Sauzet,}
\author[b]{E. Mobio,}
\author[b]{R. Babut,}
\author[c]{and C. Tao}

\affiliation[a]{Laboratoire de Physique Subatomique et de Cosmologie, Universit\'{e} Grenoble-Alpes, CNRS/IN2P3,\\38000 Grenoble, France}
\affiliation[b]{Institut de Radioprotection et de Sûreté Nucléaire (IRSN), BP 3, 13115, Saint-Paul-lez-Durance, Cedex, France}
\affiliation[c]{Centre de Physique des Particules de Marseille, Aix-Marseille Universit\'e, CNRS/IN2P3, Marseille, France}

\emailAdd{beaufort@lpsc.in2p3.fr}
\emailAdd{guillaudin@lpsc.in2p3.fr}
\emailAdd{santos@lpsc.in2p3.fr}
\emailAdd{sauzet@lpsc.in2p3.fr}

\abstract{
Directional detection is the dedicated strategy to demonstrate that DM-like signals measured by direct detectors are indeed produced by DM particles from the galactic halo. The experimental challenge of measuring the direction of DM-induced nuclear recoils with (sub-)millimeter tracks has limited, so far, the maximal directional reach to DM masses around $100~\rm{GeV}$. In this paper, we expose the MIMAC detector to three different neutron fields and we develop a method to reconstruct the direction of the neutron-induced nuclear recoils. We measure an angular resolution better than $16^\circ$ for proton recoils down to a kinetic energy of $4~\rm{keV}$ and for carbon recoils down to a kinetic energy of $5.5~\rm{keV}$. For the first time, a detector achieves the directional measurement of proton and carbon recoils with kinetic energies in the keV range without any restriction on the direction of the incoming particle. This work demonstrates that directional detection is around the corner for probing DM with masses down to $\mathcal{O}(1~\rm{GeV})$.
}

\begin{document}
\maketitle
\flushbottom

\section{Introduction}

The observation of a signal possibly attributed to dark matter (DM) in a direct detector would open an exciting and fruitful sequence in multiple research fields. However, the demonstration that such a signal is actually associated with DM is a non-trivial challenge often left to the future \cite{Billard2011, Akerib2022}. Directional detection overcomes this limitation by measuring an unambiguous DM signature \cite{Spergel1987, Vahsen2021}. In a direct detector, DM-nucleus interactions would induce nuclear recoils having an anisotropic angular distribution correlated with the motion of the Earth through the galactic DM halo \cite{Catena2015, Kavanagh2015}. Such an angular distribution cannot be reproduced by the background \cite{Mayet2016}. Directional detection consists then in measuring the directions of nuclear recoils in order to confirm, or invalidate, their correlation with the galactic DM. 

The WIMP is the archetype of a galactic DM particle that can be discovered by directional detection. For simplicity, in the following we call WIMP any galactic DM particle whose elastic scattering with a nucleus can be described by a non-relativistic effective field theory \cite{Fan2010, Fitzpatrick2012}. While the WIMP parameter space is probed by multiple direct experiments \cite{PDG2022}, the experimental constraints relax for a WIMP mass below $30~\rm{GeV}$ due to the energy thresholds of the detectors. As we will discuss in Section~\ref{sec:directional}, directional detectors are suited to explore this low-mass WIMP region for which their relatively low exposures (compared to non-directional detectors) can be compensated by the use of light targets. Besides being the appropriate strategy to discover galactic DM and being able to explore the low-mass WIMP region, directional detection is also an efficient tool to navigate through the neutrino fog which represents an irreducible background that can hardly be faced without the directional information \cite{OHare2015, Grothaus2014, OHare2022}. 

The majority of directional detectors \cite{Baracchini2023} are based on gaseous Time Projection Chambers (TPCs), such as the MIMAC detector. In gaseous directional detectors, WIMPs with masses below $30~\rm{GeV}$ would typically induce nuclear recoils with kinetic energies below $10~\rm{keV}$, corresponding to track lengths usually shorter than one millimeter. The experimental challenge consists then in measuring the directions of sub-millimiter nuclear recoils with energies in the keV range. To the best of the authors' knowledge, such directional performance was not achieved in a general configuration, we refer to \cite{OHare2022, Vahsen2020, ThesisCyprien} for some reviews. Recently, we published MIMAC measurements \cite{Beaufort2021} demonstrating a better than $15^\circ$ angular resolution on keV proton tracks produced by mono-energetic neutrons, under the experimental restriction that the incident neutrons (and by extension the WIMPs) must be aligned with the drift direction of the MIMAC detector. The current paper extends this work to a general basis, \textit{i.e.} without restriction on the direction of the incoming particle, and to keV carbon recoils. 

We begin with an overall presentation of the paper in Section~\ref{sec:overview} that briefly introduces the MIMAC detector and the measurement facilities, as well as the interest in using neutron fields to measure the directional performance of a detector. This section aims to present the analysis steps for a global understanding while letting most of the technicalities into the appendices. Section~\ref{sec:highGain} is dedicated to the description of the new method used to reconstruct the direction of the nuclear recoils. To access directionality on nuclear recoils in the keV range, the MIMAC detector must operate at high gain ($>10^4$) and with a large amplification region ($512~\rm{\mu m}$). As we will see, these operating conditions improve the sensitivity of the detector but, at the same time, distort the nuclear tracks. We develop a new method to access directionality in this context. This method is applied to three different measurements with neutron fields whose results are presented in Section~\ref{sec:measurements}. We quantify the efficiency of the new method by measuring its angular resolution in a given energy range and by evaluating its biases. The neutron measurements demonstrate the directional performance of MIMAC on keV proton and carbon recoils. Finally, Section~\ref{sec:directional} corresponds to a discussion on the possible impacts of the current work, and more generally of directional detection, in the field of direct detection of WIMPs. 

\section{Overview of the experiments and analyses}\label{sec:overview}

\begin{figure}[t]
	\centering
	\begin{minipage}{0.6\linewidth}	
		\includegraphics[width=\linewidth]{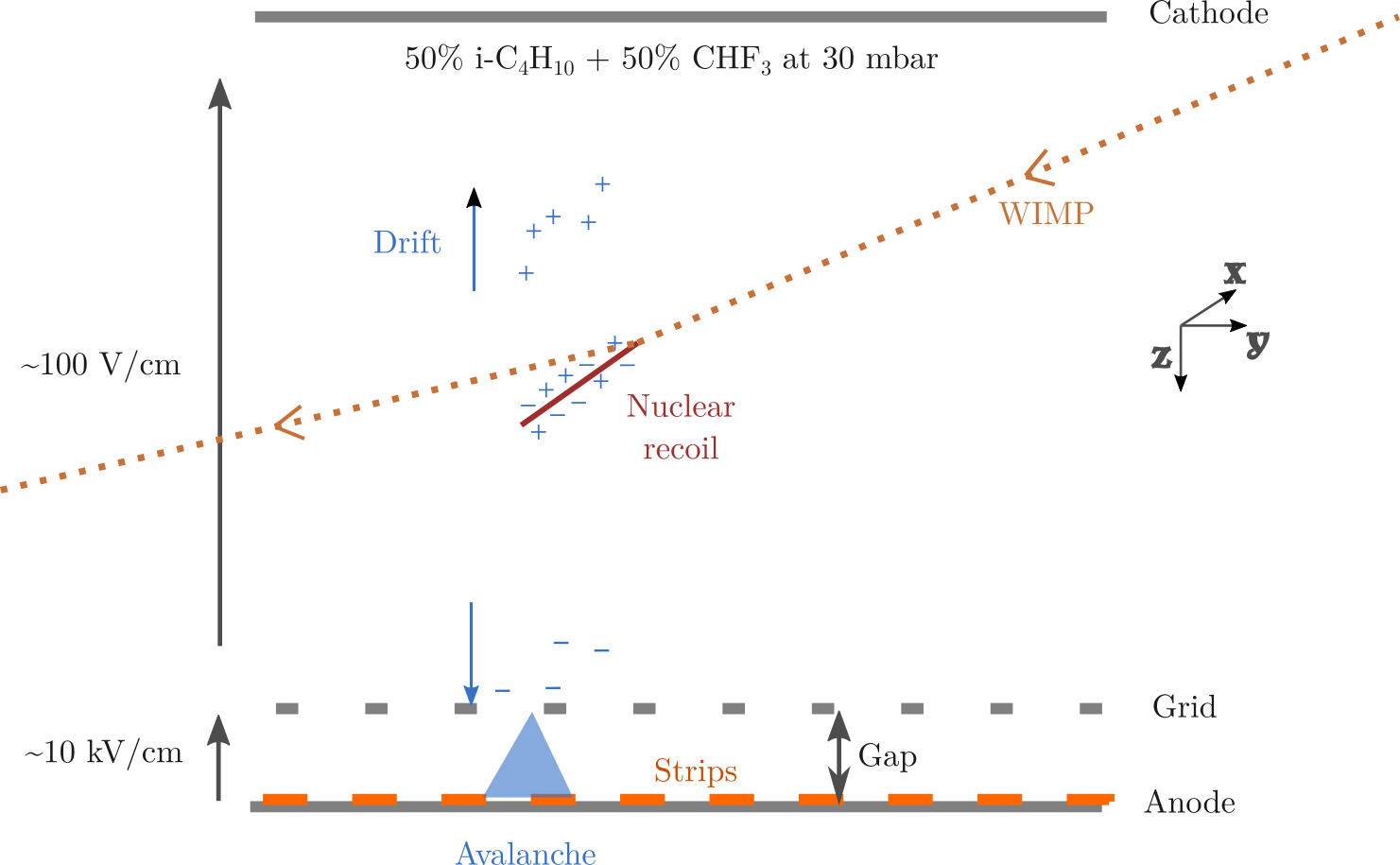}
	\end{minipage}
	\hfill
	\begin{minipage}{0.3\linewidth}	
		\includegraphics[width=\linewidth]{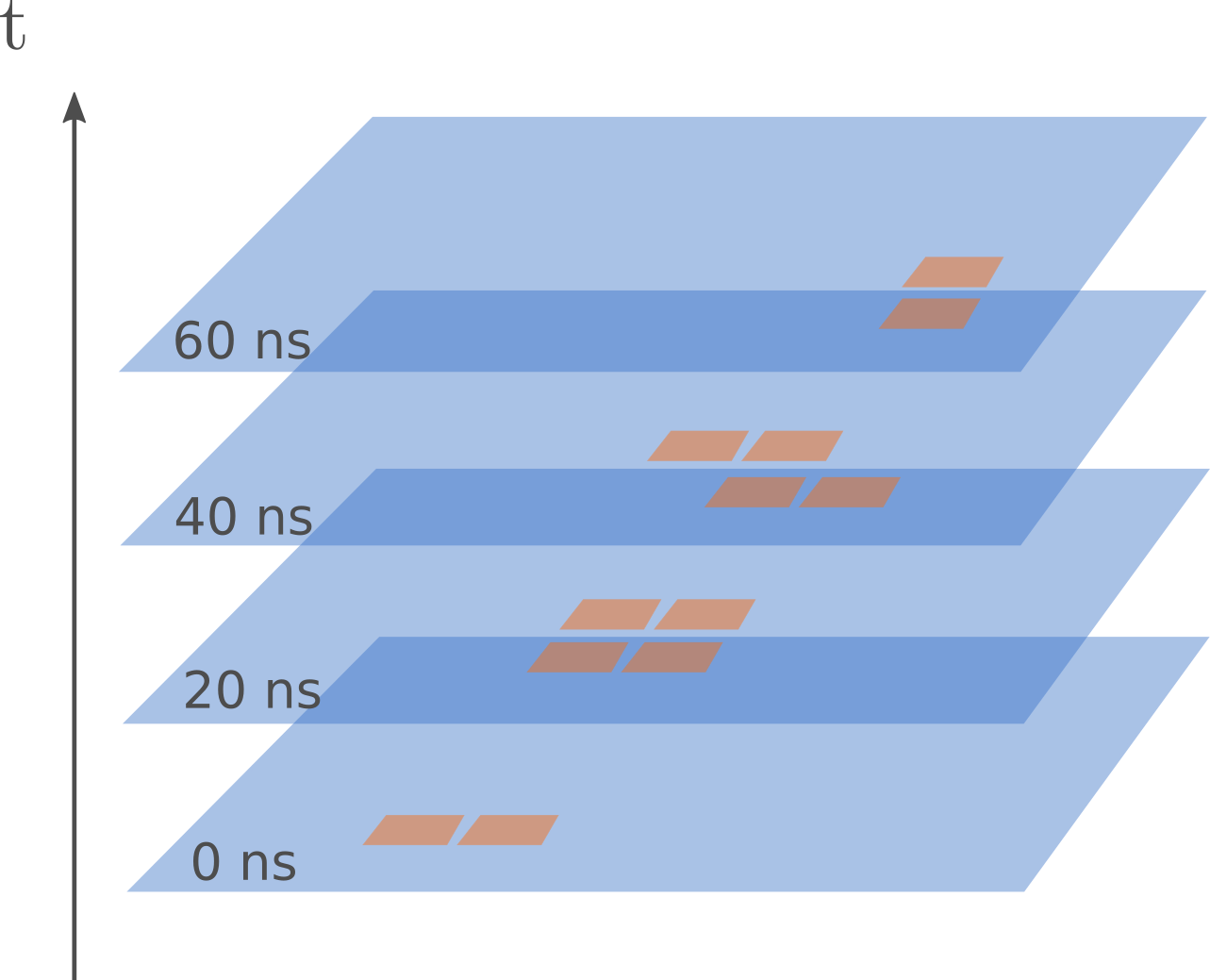}
	\end{minipage}
	\caption{Operating principles of the MIMAC detector.}
	\label{fig:MimacPrinciples}
\end{figure}

MIMAC \cite{Santos2013} is a directional gaseous detector developed at the LPSC, France, and currently operating in the Underground Laboratory of Modane. We here briefly present the detector based on the schematics of Figure~\ref{fig:MimacPrinciples} and we refer to \cite{ThesisCyprien} for a recent and comprehensive description. MIMAC relies on the principle of a Time Projection Chamber with a \textit{drift} region of $25~\rm{cm}$, based on a Micromegas with a large gap of $512~\rm{\mu m}$ for the \textit{amplification} region. The detector is filled with a mixture of $50\%$ i-C$_4$H$_{10}$ + $50\%$ CHF$_3$ at $30~\rm{mbar}$ dedicated to keV recoils detection \cite{Beaufort2021} and mainly composed of odd nuclei (H and F) opening the spin-dependent channel for WIMP detection \cite{Cerdeno2011}. The primary electrons, resulting from the ionization process by a nuclear recoil, drift towards the anode under an electric field of $84~\rm{V/cm}$ leading to a drift velocity $v_{\rm{drift}} = 11.5~\rm{\mu m/ns}$ according to the \texttt{Magboltz} simulation code \cite{Biagi1999}. When a primary electron enters the amplification region, it produces an avalanche leading to more than $10^4$ secondary electrons and ions. The secondary ions are collected on the grid of the Micromegas whereas the secondary electrons are collected on the anode. The ionization energy is measured on the grid by a Flash Analog to Digital Converter, herefater denoted \textit{Flash-ADC}. The track is reconstructed from the pixelated anode, having $256$ strips in each direction, enabling a 2D measurement of the positions of the charge. The signal is read every $20~\rm{ns}$ and the constant drift velocity allows the reconstruction of the longitudinal direction, leading to the 3D track reconstruction.

In this paper, we present measurements performed with two facilities:
\begin{itemize}
	\item The Comimac facility \cite{Muraz2016} is a table-top accelerator in which a gas mixture in an Electron Cyclotron Resonance Ion Source (ECRIS) is excited by 2.5 GHz radio-frequency waves generating a plasma. A voltage is applied to extract electrons or ions of known kinetic energy from a few hundred of eV up to $50~\rm{keV}$. By interfacing Comimac with a MIMAC chamber, we have studied in detail the signal formation in the detector \cite{Tao2019, Tao2020, Beaufort2021}. In this work, we use Comimac to calibrate the detector and to measure the Ionization Quenching Factor. 
	\item The AMANDE facility \cite{Gressier2008, Gressier2014} of the French Institute for Radiation protection and Nuclear Safety (IRSN) is a neutron metrologic installation. Protons or deuterons beams are sent to thin targets to produce nuclear reactions generating mono-energetic neutron fields. In this work, we analyze two kinds of measurements performed on this facility with the MIMAC detector:
	\begin{itemize}
		\item Mono-energetic neutron fields obtained from resonances of the $^{45}$Sc$(p,n)^{45}$Ti nuclear reaction. We analyze measurements performed in 2020 with a $27.24\pm 0.05~\rm{keV}$ field and measurements of 2021 with an $8.12\pm 0.01~\rm{keV}$ field. We published a first analysis of these measurements in \cite{Beaufort2021}.
		\item A bi-energetic neutron field from the nuclear reaction $^7$Li$(p,n)^7$Be near threshold that generates a neutron spectrum with a mean energy of about $30~\rm{keV}$  \cite{Martin-Hernandez2016, Lee1999, Herrera2015}. In the experiment, we use a proton energy of $1881\pm1~\rm{keV}$ which is about $1~\rm{keV}$ above the threshold. The proton beam is sent on a LiF target of thickness $117~\rm{\mu g/cm^2}$.
	\end{itemize}
\end{itemize}

The AMANDE facility represents a valuable tool for directional detection because neutrons induce a similar signal than WIMPs. The elastic scattering of a neutron $n$ with a nucleus of mass $m_R$ of the gas mixture produces a nuclear recoil with a kinetic energy $E_R$ related to the neutron kinetic energy $E_n$ via the scattering angle $\theta$:
\begin{equation}
	E_R = \frac{4 m_n m_R}{(m_n +m_R)^2}\, E_n\,\cos^2\theta\,.
	\label{eq:kinematics}
\end{equation}

The simultaneous measurement of $E_R$ and $\theta$, which are the relevant observables for a directional detector, leads to the reconstruction of the neutron energy spectrum. By comparing the reconstructed neutron energy with the expected one, we evaluate the performance of a directional detector such as MIMAC.

While most of the technicalities of the analyses are left to the appendices, we here present the general guidelines to reconstruct the neutron spectrum and evaluate the directional performance of the detector:
\begin{enumerate}
	\item Energy reconstruction. The MIMAC detector measures the energy deposited in ionization in the gas mixture based on the calibration of the Flash-ADC (\textit{c.f.} Appendix~\ref{app:calib}). For nuclear recoils, the ionization energy is then converted into a kinetic energy by means of the Ionization Quenching Factor (\textit{c.f.} Appendix~\ref{app:IQF}). In the following, we note IQF the Ionization Quenching Factor.
	\item Electron-recoil discrimination. The neutron production in AMANDE comes with a large production of gamma-rays that generate electron-like signals in the detector. For instance, for the $27~\rm{keV}$ neutron field, only $0.1\%$ of the measured events are identified as nuclear recoils. As detailed in Appendix~\ref{app:eRdiscri}, we make use of two datasets for each campaign: one "background only" and one "background + signal". For the $27~\rm{keV}$ and the $8~\rm{keV}$ measurements, the discrimination electron-recoil is performed by means of Boosted Decisions Trees (BDTs) whereas for the bi-energetic neutron field of 2023 a discrimination based on minimal cuts turned out to be sufficient because of lower production of gamma-rays.
	\item Proton-carbon discrimination. For a low enough energy threshold, we expect to detect both proton and carbon recoils in the i-C$_4$H$_{10}$ + $50\%$ CHF$_3$ mixture. For instance, a $27~\rm{keV}$ neutron can induce a carbon recoil with a maximal kinetic energy of $7.35~\rm{keV}$ according to Eq.~\eqref{eq:kinematics}. For the bi-energetic measurement, a few tens of fluorine recoils also exceed the threshold but we will see that we do not discriminate them from the carbon recoils. As detailed in Appendix~\ref{app:pCdiscri}, we separate the carbon from the proton recoils by using BDTs. 
	\item Angle reconstruction. The reconstruction of the angle of the nuclear recoil, leading to the elastic scattering angle reconstruction, requires a non-trivial process that is presented in the next section. 
\end{enumerate}

We provide at this stage a summary of the analyses in Table~\ref{tab:summary} for a better understanding of the paper. Since the new angular reconstruction method proposed in this work relies on parameters fixed experimentally, the consistency between all measurements is a key element for evaluating the robustness of the method. We emphasize that the measurements are performed at three different high voltages applied on the grid, corresponding to three different gains for the avalanches. However, the electric field in the drift region (and thus the drift velocity) is kept constant during all the measurements. 

\begin{table}[t]
\hspace*{-1cm}
\begin{tabular}{ccccccc}
\hline \hline
\multirow{3}{*}{Run}            & \multirow{3}{*}{Year} & \multirow{3}{*}{\begin{tabular}[c]{@{}c@{}}High voltage\\on the grid\end{tabular}} & \multicolumn{2}{c}{\begin{tabular}[c]{@{}c@{}}Kinetic energy range\\of directional detection\end{tabular}} & \multicolumn{2}{c}{\begin{tabular}[c]{@{}c@{}}Angular resolution\end{tabular}} \\
\\
                                &                       &                               & Proton                                                   & Carbon                                                  & Proton                                                        & Carbon                                                       \\ \hline
Mono-energetic 27 keV           & 2020                  & 510V                          & $[7,~20~\rm{keV}]$                                       & $[5.5,~7.5~\rm{keV}]$                                     & $10^\circ$                                                    & $15^\circ$                                                   \\
Mono-energetic 8 keV            & 2021                  & 530V                          & $[4,~8~\rm{keV}]$                                        & /                                                       & $16^\circ$                                                    & /                                                            \\
Bi-energetic & 2023 & 520V         & $[2,~40~\rm{keV}]$                      & $[5,~15~\rm{keV}]$                    &                                            / & /                      \\ \hline \hline
\end{tabular}
\caption{Summary of the data analyses presented in this paper. We give the range for which we have been able to reconstruct the direction of the nuclear recoils, both for protons and carbons. Due to the energy threshold, no carbon recoil is expected for the mono-energetic $8~\rm{keV}$ campaign. As detailed in Section~\ref{sec:measurements}, the angular resolution can only be determined from a mono-energetic neutron field, which explains the lack of data in the last raw.}
\label{tab:summary}
\end{table}

\section{Angle reconstruction at high gain and large gap}\label{sec:highGain}

The angle reconstruction with MIMAC from the pixelated anode has previously been experimentally validated for intermediate gains ($\lesssim  10^4$) and an amplification gap of $256~\rm{\mu m}$ \cite{Maire2013, Maire2016}. However, such a detector configuration cannot access directionality for a WIMP mass typically below $30~\rm{GeV}$ that produces recoils in the keV range with track lengths of the order of the millimeter, or even less. To explore the low-mass WIMP region with directionality, we take advantage of what appears, at first glance, as a contradiction when operating at high gain $(>10^4$) and a large gap of $512~\rm{\mu m}$: the signal is amplified so the detector gets more sensitive while the 3D tracks suffer from distortion.

Let us briefly introduce the situation and we refer to chapter 8 of \cite{ThesisCyprien} for a detailed description. The current induced both on the grid and on the strips of the anode is proportional to the number and the velocities of the secondary charges (ions and electrons) located in the amplification region. The electrons are collected in about $1~\rm{ns}$ compared to a few hundred of nanoseconds for the ions. The larger the gain, the more secondary charges produced in the avalanches. The larger the gap, the longer it takes for the ions to be collected on the grid, and therefore the longer the induced signal. At high gain and large gap, millions of secondary ions become accumulated in the amplification region. These ions induce a baseline on the signal which improves the sensitivity of the detector but, on the other hand, results in an elongation of the measured track since a signal is still detectable about $300~\rm{ns}$ after the avalanche produced by the last primary electron. 

The Flash-ADC measures the signal induced on the grid which corresponds to the charge integration as a function of time. In \cite{Beaufort2021}, we developed an analytical deconvolution formula to separate the electronic signal from the ionic signal measured by the Flash-ADC without introducing any \textit{ad hoc} parameter. Since each avalanche induces an electronic signal for a period of about $1~\rm{ns}$, which is lower than the MIMAC time resolution, the extracted electronic signal provides the time distribution of the primary electrons cloud at the grid level, \textit{i.e.} before the signal distortion by the ionic component. We then proposed a method to determine the nuclear recoil direction by exploiting this deconvolution although, as already mentioned, our procedure requiring the incident particle to be aligned with the drift direction of the detector, this condition must be taken into account in the analysis of WIMP directional detection.

The pixelated anode corresponds to the part of the MIMAC detector initially dedicated to the measurement of the directionality. Developing a mathematical deconvolution similar to the one established for the Flash-ADC is limited by some specificities of the signal formation on the anode's strips: the weighting field used to determine the induced current is non-linear; the strips only provide binary information: fired or not; they have thresholds; and they are coupled to a preamplifier acting as an RC circuit. These specificities make impossible the development of a mathematical deconvolution similar to the one applied to the Flash-ADC. Therefore, we propose a new method, called the \textit{high gain 3D method}, to reconstruct the angle from the pixelated anode at high gain and large gap.

\begin{figure}[t]
	\centering
	\includegraphics[width=\linewidth]{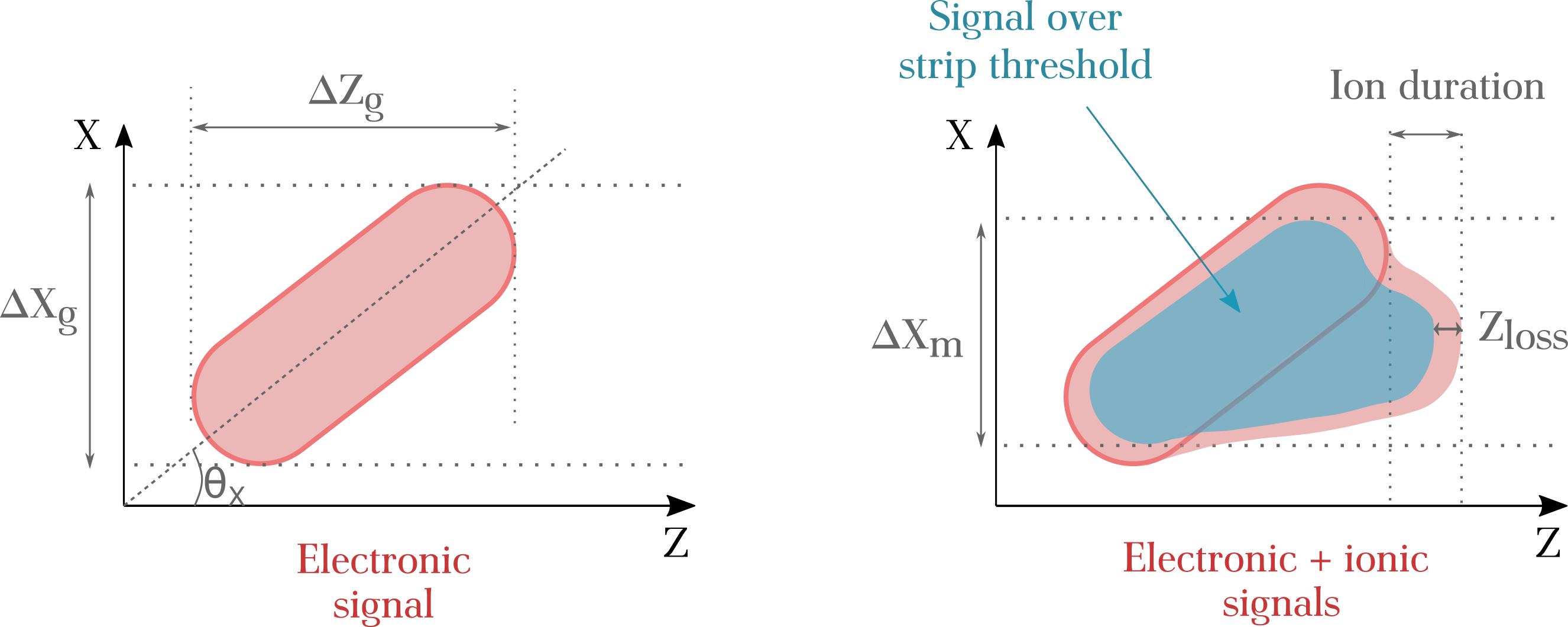}
	\caption{2D representation of a track measurement on the anode. The figure on the left sketches the situation with only the electronic signal represented whereas the right figure also includes the ionic signal. The blue-shaded region corresponds to the signal that exceeds the strip threshold, \textit{i.e.} the anode measurement. The quantities used to access the track directionality from the anode are shown in the figure.}
	\label{fig:anodeMethod}
\end{figure}

Let us describe the high gain 3D method by using Figure~\ref{fig:anodeMethod} as an illustration. Instead of reconstructing the angle from the 3D track measurement, we project the measurement in the (X-Z) and (Y-Z) planes to determine the angles $\theta_X$ and $\theta_Y$ related to the polar angle $\theta$ and the azimuthal angle $\phi$ by:
\begin{equation}
	\tan\theta_X = \tan\theta \, \cos \phi \hspace*{2cm} \tan\theta_Y = \tan\theta \, \sin\phi~.
	\label{eq:polarAzimuthal}
\end{equation}

As illustrated in the left panel of Figure~\ref{fig:anodeMethod}, the $\theta_X$ angle at the level of the grid (and similarly for $\theta_Y$) is given by:
\begin{equation}
	\tan \theta_X = \frac{\Delta X_g}{\Delta Z_g} \, \frac{D_L}{D_T}~,
\end{equation}
in which $\Delta X_g$  and $\Delta Z_g$ correspond to the width and the length, respectively, of the primary electrons cloud at the grid level, \textit{i.e.} before the avalanches. The ratio of the longitudinal to the transverse diffusion coefficients, $D_L$ over $D_T$, accounts for the diffusion during the drift. These coefficients can be determined by the \texttt{Magboltz} simulation code \cite{Biagi1999}. The value of $\Delta Z_g$ can be determined from the deconvolution of the Flash-ADC since it gives access to the time distribution of the primary electrons cloud. 

However, the quantity $\Delta X_g$ cannot be inferred directly from the anode measurement since it is affected by the ionic signal and the way the strips are fired, as previously mentioned, resulting in a distortion of the track measurement such as represented in blue in the right panel of Figure~\ref{fig:anodeMethod}. We propose an estimation of $\Delta X_g$ from the width measured on the anode, noted $\Delta X_m$, to which a corrective factor is applied:

\begin{equation}
	\tan \theta_X ~=~ \frac{\Delta X_g}{\Delta Z_g} \, \frac{D_L}{D_T} ~\simeq~ \frac{\Delta X_m}{\Delta Z_g} \, \frac{D_L}{D_T}   ~  \bigg(\gamma \frac{\rm{Z_{loss}}}{\rm{Ion~duration}}\bigg)^\delta~.
\end{equation}

The quantity \texttt{$\rm{Z_{loss}}$}, represented in Figure~\ref{fig:anodeMethod}, corresponds to the duration during which the Flash-ADC continues to detect a signal on the grid whereas the anode strips are no longer fired. The quantity \texttt{Ion duration} is the duration of the measured Flash-ADC signal once the last primary electron has reached the grid. In other words, it is the duration of the measured ionic signal. The form of the corrective factor is motivated by two elements: using the additional information provided by the Flash-ADC along the Z-axis to estimate the loss along the X-axis; using the ratio of two quantities similarly affected by the ionic signal in order to provide a robust correction. The parameters $\gamma$ and $\delta$ are introduced to make compatible these elements.

\begin{figure}[t]
	\centering
	\begin{minipage}{0.49\linewidth}
		\includegraphics[width=\linewidth, height=6cm]{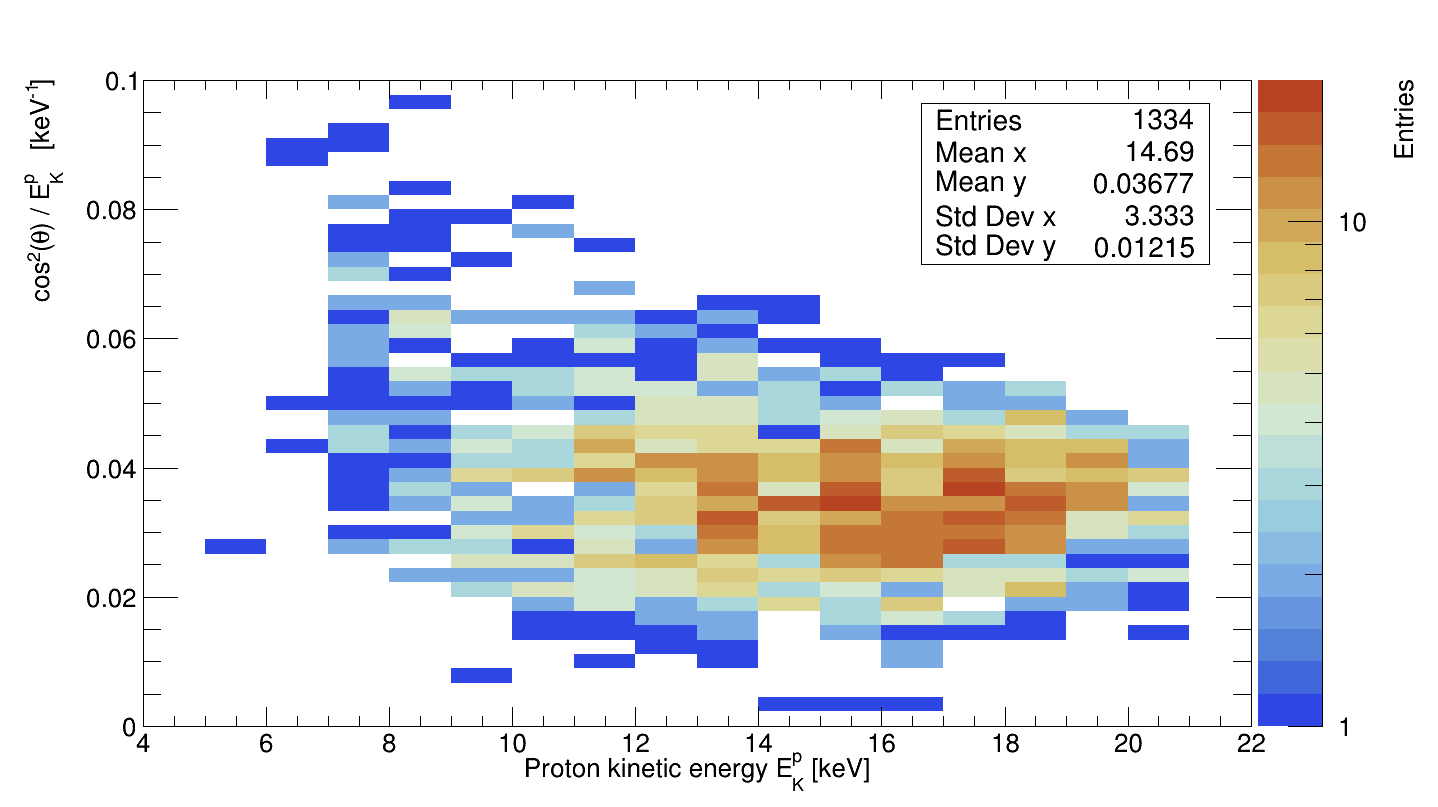}
	\end{minipage}
	\hfill
	\begin{minipage}{0.49\linewidth}
		\includegraphics[width=\linewidth, height=6cm]{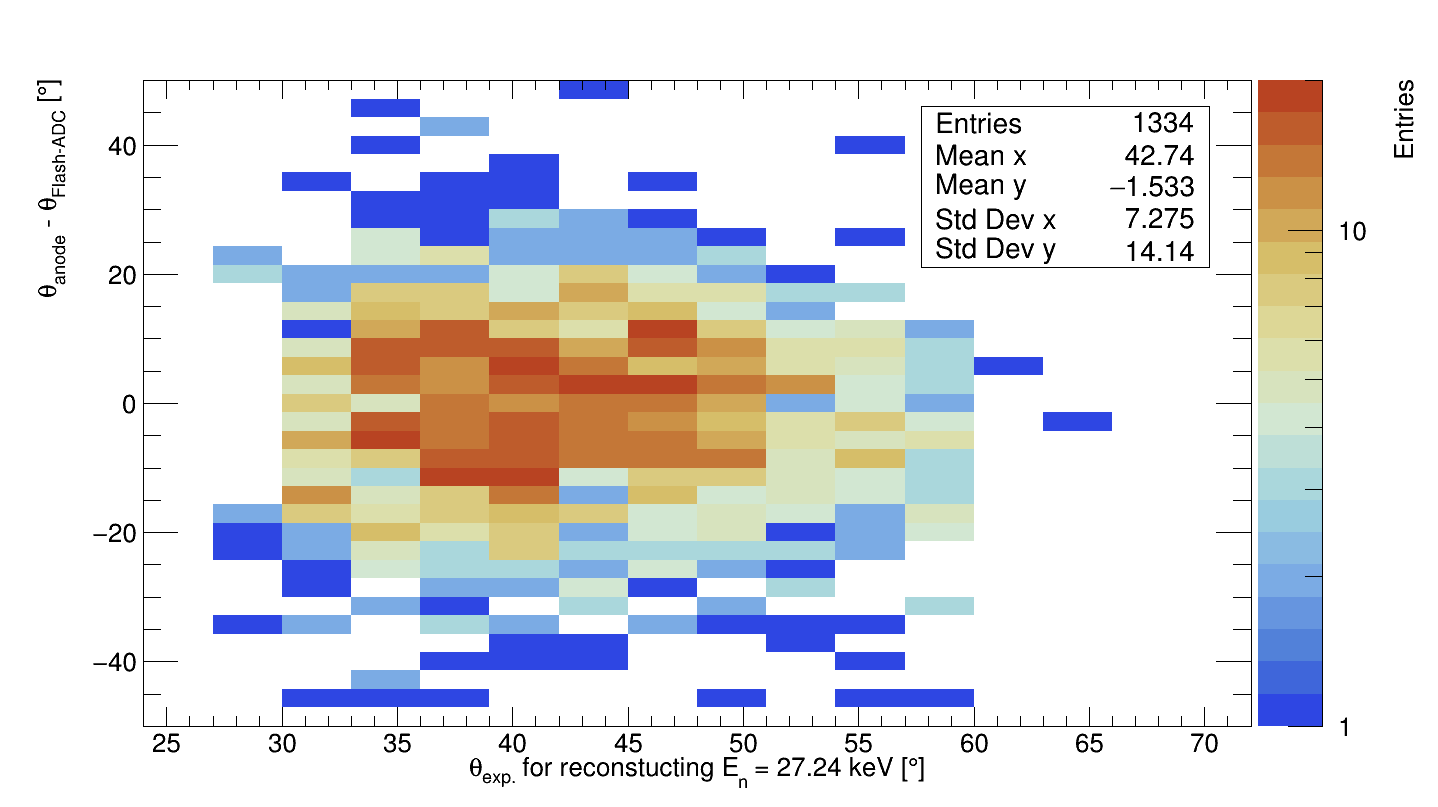}
	\end{minipage}
	\caption{Evaluation of the high gain 3D method for the parameters $\gamma = 5/3$  and $\delta = 5/4$ in the mono-energetic $27~\rm{keV}$ neutron field. Left: the ratio $\cos^2(\theta) / E_K^p$ as a function of the proton kinetic energy. Right: the difference between $\theta_ {\rm{anode}}$ and $\theta_ {\rm{Flash-ADC}}$, respectively the angle reconstructed from the high gain 3D method and the angle reconstructed from the Flash-ADC method of \cite{Beaufort2021}, as a function of the expected angle to reconstruct the incoming neutron energy.}
	\label{fig:fixingParameters}
\end{figure}

The two parameters are fixed by using a reference dataset from a mono-energetic neutron field. In this paper, we have chosen the proton recoils from the $27~\rm{keV}$ field since, for this configuration, the directional reconstruction from the Flash-ADC in \cite{Beaufort2021} has already validated the calibration, the IQF, and the electron-recoil discrimination. Due to the mono-energetic field, we expect the value of $\cos^2\theta / E_K^p$ to be constant and equal to $1 / E_n$, where $E_K^p$ and $E_n$ are the kinetic energies of the proton recoil and of the neutron, respectively. Moreover, the angular distribution of proton recoils resulting from elastic scatterings with neutrons is centered on $45^\circ$, as attested by a GEANT4 simulation. Those two requirements lead to a mean estimation of the parameters to $\gamma = 5/3$  and $\delta = 5/4$. The uncertainties on these parameters have not been estimated. The left panel of Figure~\ref{fig:fixingParameters} shows the ratio $\cos^2\theta / E_K^p$ obtained with such parameters that, except for some dispersion, follows the expected plateau centered on $1/E_n = 0.037~\rm{keV^{-1}}$ for the entire energy range of the proton recoils. 

Several comments are required at this stage:
\begin{itemize}
	\item While this method relies on experimentally fixed parameters, we will see that it leads to consistent results with the three datasets considered at multiple energies and gain, and equally for proton or carbon recoils. It requires a reference measurement with a mono-energetic neutron field to calibrate the $\gamma$ and $\delta$ parameters. This reference measurement must be repeated for each gas mixture considered. It may possibly be repeated for each value of drift electric field since, in our work, we have not varied this parameter so we cannot conclude on its influence.
	\item The quantities \texttt{$\rm{Z_{loss}}$} and \texttt{Ion duration} are both affected by the way the strips are fired, the gain, and the density of the primary charges. Their ratio provides a robust corrective factor that is adapted to multiple experimental conditions. 
	\item The larger the angle, the more difficult its reconstruction, since its track is shorter and is less amplified by the ionic baseline because the transverse distance (perpendicular to the anode) between two consecutive avalanches is larger. Both effects limit the resolution for defining \texttt{$\rm{Z_{loss}}$} and \texttt{Ion duration} and are mainly observed for angles larger than $45^\circ$. We thus apply a validity condition by requiring that the density of activated pixels on the anode is large enough to correctly reconstruct the angle. This validity condition removes on average $25\%$ of the recoil events. 
	\item The right panel of Figure~\ref{fig:fixingParameters} compares the reconstructed angles from the high gain 3D method and from the Flash-ADC method previously introduced in \citep{Beaufort2021}. The difference between the two reconstructed angles has an offset of $1.5^\circ$ and a dispersion of $\sigma=14^\circ$. These discrepancies are comparable to the angular resolution of both methods. For this reason, and since the difference between the angles does not show any dependency on the expected angle, we conclude that the two methods are consistent.
\end{itemize}

\section{Directional performance measurements with neutron fields}\label{sec:measurements}

This section presents the main directional results obtained with the high gain 3D method to reconstruct the scattering angle from the anode for recoils in the keV range. The high gain 3D method reconstructs the polar and azimuthal angles from Eq.~\eqref{eq:polarAzimuthal}. The neutron-recoil scattering angle is then obtained by accounting for the angle of the incident neutron before the elastic collision. To determine such an angle, we know the distance between the neutron source and the detector and we know from the anode the 2D coordinates of the nuclear recoil emission point. The only missing information is the absolute Z-position (parallel to the drift direction) of the interaction inside the active volume. We assume the interaction to happen at the center of the active volume in the Z-axis, leading to a maximal error of $0.5\times25~\rm{cm}$ in Z corresponding to $1.7^\circ$ on the reconstruction of the neutron-recoil scattering angle once accounting for the distance between the neutron source and the detector.

\subsection{Mono-energetic 27 keV field}\label{subsec:27keV}

\begin{figure}[t]
	\centering
	\begin{minipage}{0.49\linewidth}
		\includegraphics[width=\linewidth, height=6cm]{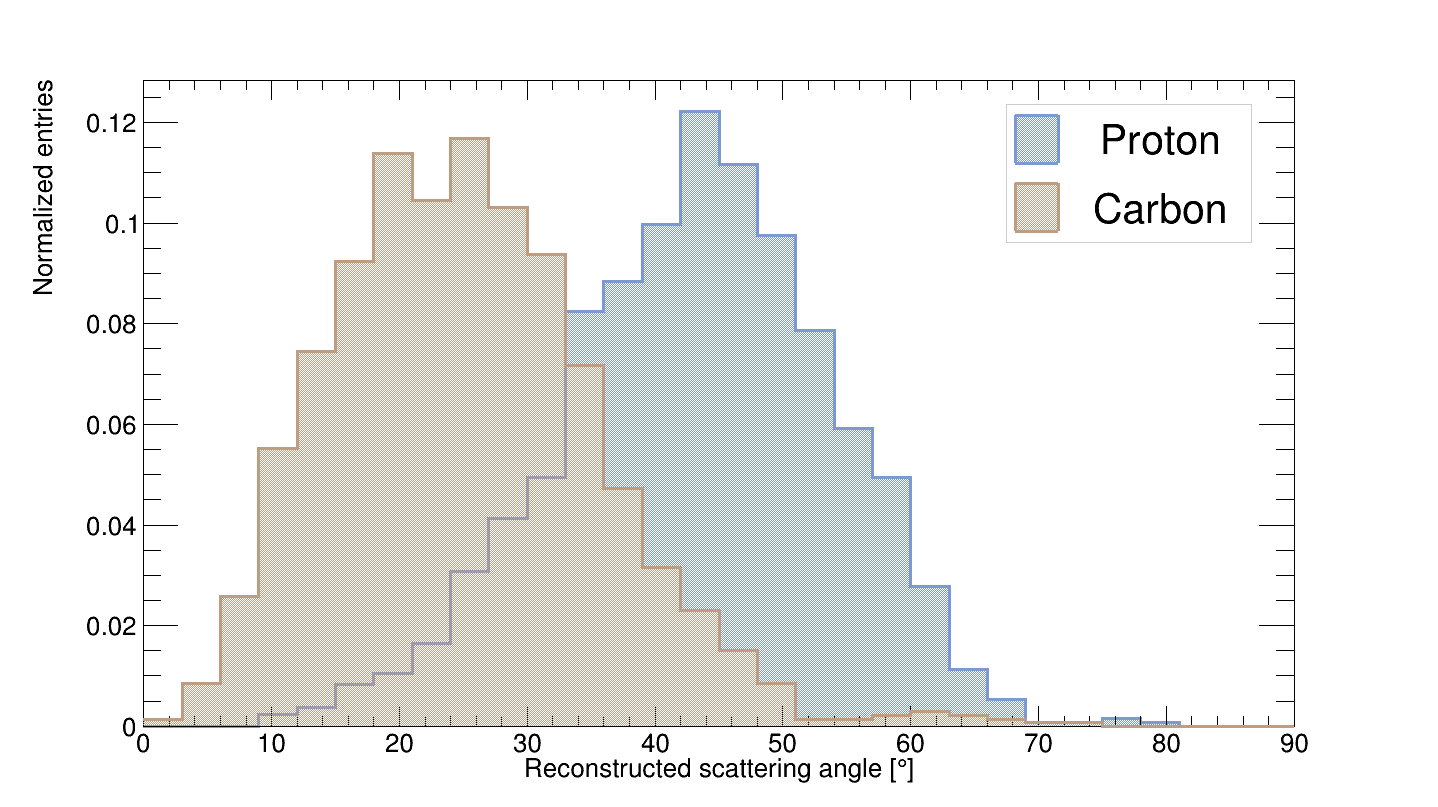}
	\end{minipage}
	\hfill
	\begin{minipage}{0.49\linewidth}
		\includegraphics[width=\linewidth, height=6cm]{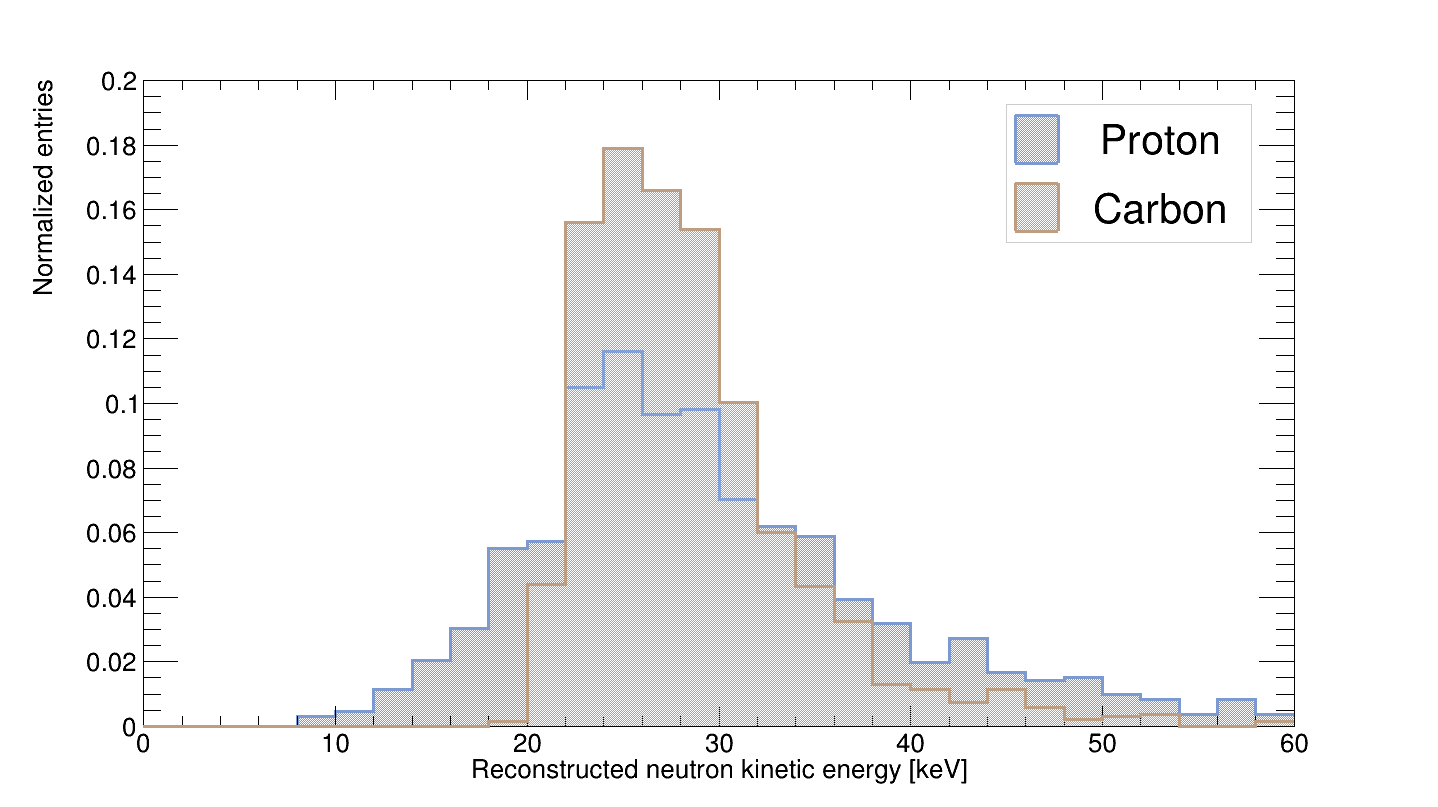}
	\end{minipage}
	\caption{Angular distribution (left) and neutron energy spectrum (right) reconstructed from the proton and carbon recoils detected in the $27~\rm{keV}$ campaign. The scattering angle is reconstructed from the high gain 3D method. The neutron energy spectrum embeds all the systematic uncertainties of the analysis (calibration, IQF, discriminations, angle reconstruction). The figures are constructed from $1334$ proton recoils and $1429$ carbon recoils.}
	\label{fig:27keV_spectrumAngle}
    \vspace*{0.1cm}
	\centering
	\includegraphics[width=\linewidth]{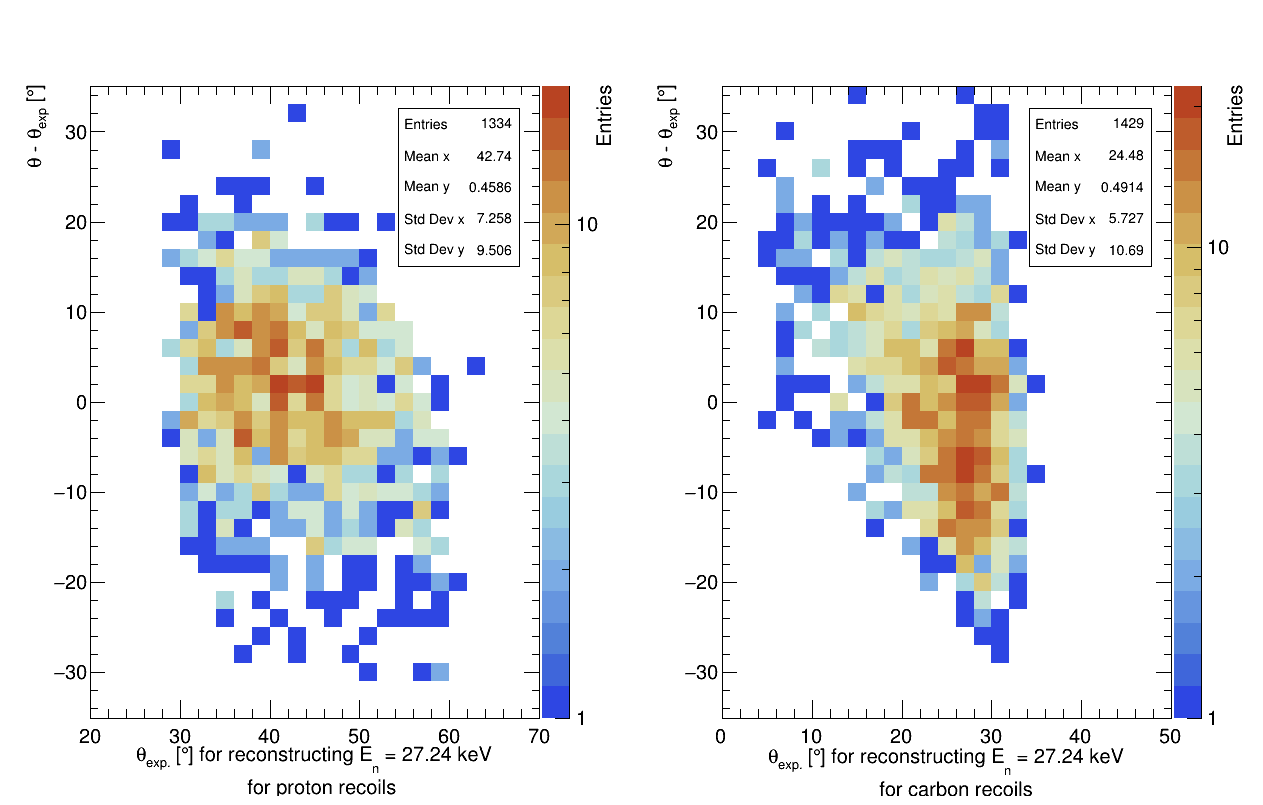}
	\caption{Difference between $\theta$ the scattering angle reconstructed from the high gain 3D method and $\theta_{\rm{exp.}}$ the expected angle to reconstruct the neutron energy of $27.24~\rm{keV}$, as a function of $\theta_{\rm{exp.}}$. Left: for the proton recoils. Right: for the carbon recoils.}
	\label{fig:27keV_thetaExp}
\end{figure}

Figure~\ref{fig:27keV_spectrumAngle} presents the angular distribution and the neutron energy spectrum obtained from the proton and the carbon recoils, after the analysis steps detailed in the appendices, for the $27~\rm{keV}$ campaign.

Let us first comment on the angular distribution reconstructed from the high gain 3D method. We observe that the proton distribution is centered at $45^\circ$ which is expected since it is used to calibrate the method, as explained in the previous section. However, the angular distribution for the carbon recoils covers a lower angular range. The reason for this lies in the energy range. During the campaign, we cover an ionization energy range of $[1.6~\rm{keV}, 14~\rm{keV}]$. From the kinematics Eq.~\eqref{eq:kinematics} and the IQF described in Appendix~\ref{app:IQF}, the minimum and the maximum angles of the proton and carbon recoils can be calculated, respectively, from the upper limit in ionization energy and from the lower one. Consequently, the angular range for the proton recoils is $[30^\circ, 70^\circ]$ and is $[0^\circ, 38^\circ]$ for the carbon recoils. The reconstructed angular distributions of Figure~\ref{fig:27keV_spectrumAngle} are then consistent with the expectations. They also show an overlap of some recoils identified as carbons with angles above $38^\circ$ and some recoils identified as protons below $30^\circ$. Such an overlap is likely due to the angular resolution that spreads the distributions and to some misidentifications in the proton-carbon discrimination. 

The scattering angle is then combined with the ionization energy corrected from the IQF to reconstruct the neutron energy spectrum presented in the right panel of Figure~\ref{fig:27keV_spectrumAngle}. We emphasize that such a spectrum embeds all the systematics of the analysis: the uncertainties in the high gain 3D method, but also uncertainties in the calibration, the IQF, and misidentifications in the electron-recoil and proton-carbon discriminations. Once again, due to the calibration of the high gain 3D method, the neutron spectrum reconstructed from the proton recoils corresponds to the expectation. Interestingly, the neutron spectrum reconstructed from the carbon recoils, using the $\gamma$ and $\delta$ parameters estimated from the proton data, is consistent with the reference one reconstructed with the proton recoils: it peaks at an energy $5\%$ lower than the expected value, a difference that can be considered as small regarding the multiple sources of systematic uncertainties. We emphasize that the carbon recoils have a maximal kinetic energy of $8~\rm{keV}$ (determined from the kinematics Eq.~\eqref{eq:kinematics}), meaning that the spectrum of Figure~\ref{fig:27keV_spectrumAngle} is, to the best of the authors' knowledge, the first neutron spectrum reconstruction from carbon recoils having kinetic energies in the keV range.

Let us now evaluate the resolution of the angle reconstruction from the high gain 3D method. Since this campaign is based on mono-energetic neutrons of controlled energy, we can determine the scattering angle needed to reconstruct a neutron energy of $27~\rm{keV}$ from the measurement of the ionization energy and from the IQF. We present in Figure~\ref{fig:27keV_thetaExp} the difference between this expected angle, noted $\theta_{\rm{exp.}}$, and the reconstructed angle from the anode, as a function of $\theta_{\rm{exp.}}$ both for the proton and the carbon recoils. In both cases, the difference between the angles is centered near $0^\circ$, which means that the angle reconstruction does not have a constant bias. While the difference remains relatively constant over the entire angular range for proton recoils, its mean value for carbon recoils varies with the expected angle. For carbon recoils with $\theta_{\rm{exp.}} > 15^\circ$, the angle difference is compatible with zero when accounting for the statistical uncertainties. However, this is not the case for lower angles with indicates that the high gain 3D method suffers from a bias when reconstructing the direction of carbon recoils with angles smaller than $15^\circ$. Since the majority of the detected carbon events in this analysis exceed such an angle, we anyway conclude from Figure~\ref{fig:27keV_thetaExp} that the angle reconstruction from carbon recoils is qualitatively consistent with the reconstruction from the proton recoils.

We define the angular resolution of the high gain 3D method as the standard deviation of the difference $\theta - \theta_{\rm{exp.}}$. Technically, such an angular resolution can vary over the angular range, so it might be evaluated for each bin of expected angle in the histograms of Figure~\ref{fig:27keV_thetaExp}. For simplicity, we consider an overall angular resolution that is chosen conservatively: the standard deviation remains below this value for each bin. We measure an angular resolution of $10^\circ$ for the proton recoils with kinetic energies in the range $[7, 20~\rm{keV}]$, and an angular resolution of $15^\circ$ for carbon recoils with kinetic energies in the range $[5.5, 7.5~\rm{keV}]$.

\subsection{Mono-energetic 8 keV field}\label{subsec:8keV}

Let us now apply the high gain 3D method to the measurements of the $8~\rm{keV}$ neutron data. Note that we use the values of $\gamma$ and $\delta$ parameters that have been previously estimated on the protons recoils of the $27~\rm{keV}$ experiment. No carbon recoil is expected in the measurements since the maximal energy transferred to a carbon recoil by an $8~\rm{keV}$ neutron is similar to the threshold ($460~\rm{eV}$ in ionization) of this experimental campaign. 

\begin{figure}[t]
	\centering
	\begin{minipage}{0.49\linewidth}
		\includegraphics[width=\linewidth, height=6cm]{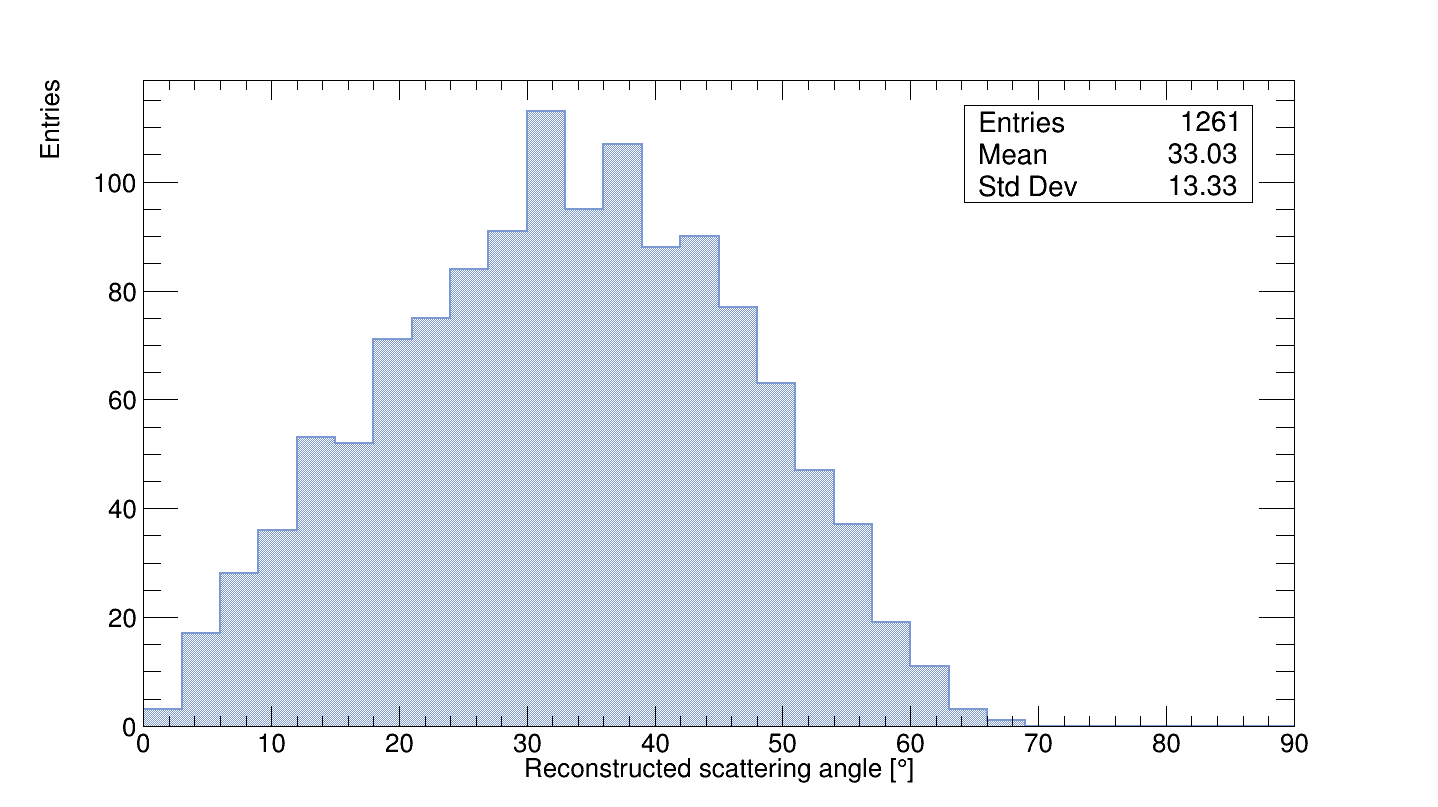}
	\end{minipage}
	\hfill
	\begin{minipage}{0.49\linewidth}
		\includegraphics[width=\linewidth, height=6cm]{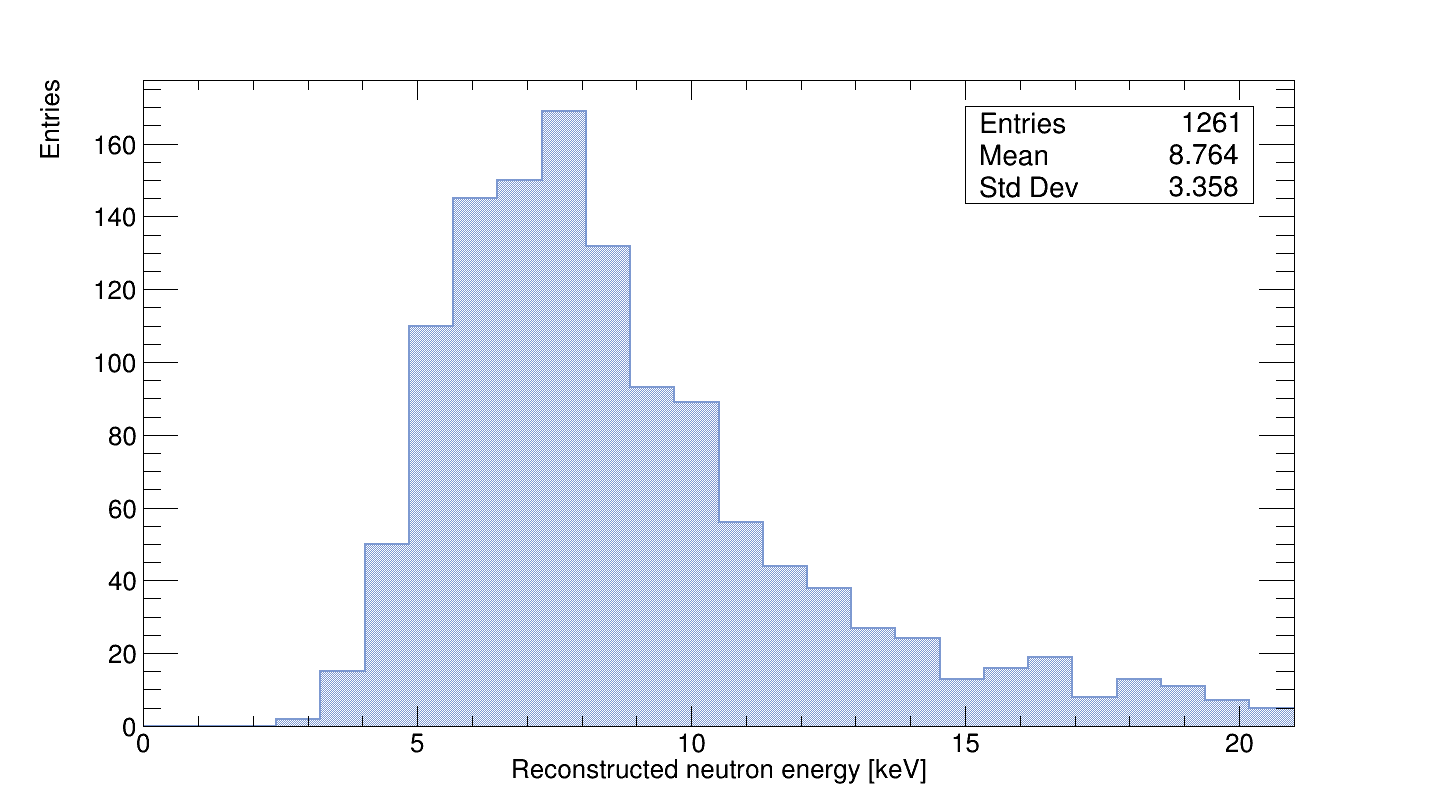}
	\end{minipage}
	\caption{Angular distribution (left) and neutron energy spectrum (right) reconstructed from the proton recoils detected in the $8~\rm{keV}$ campaign.}
	\label{fig:8keV_spectrumAngle}
    \vspace*{0.1cm}
    \includegraphics[width=\linewidth]{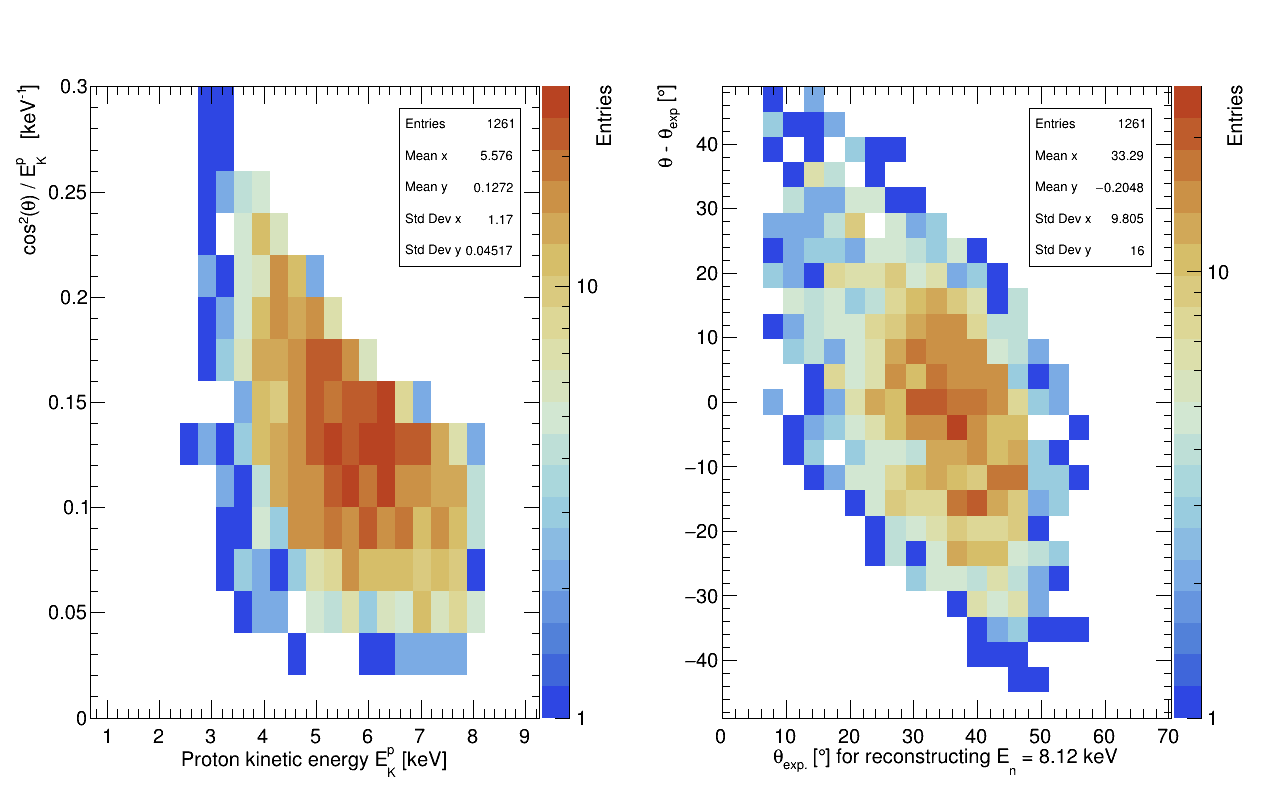}
	\caption{Evaluation of the resolution and the bias of the reconstruction of the neutron-proton scattering angle in the $8~\rm{keV}$ campaign. Left: the ratio $\cos^2(\theta) / E_K^p$ as a function of the proton kinetic energy. Right: the difference between $\theta$ the neutron-proton scattering angle reconstructed from the high gain 3D method and $\theta_{\rm{exp.}}$ the expected angle to reconstruct the neutron energy of $8.12~\rm{keV}$, as a function of $\theta_{\rm{exp.}}$.}
	\label{fig:8keV_thetaExp_cosSquared}
\end{figure}

Figure~\ref{fig:8keV_spectrumAngle} presents the angle reconstructed from the high gain 3D method and the corresponding neutron energy spectrum. As explained in Appendix~\ref{app:eRdiscri}, the electron-recoil discrimination at low energy is challenging so we apply a stringent discrimination requirement to reduce the proportion of background events passing the cut. It results in a loss of proton recoil events with kinetic energies below $5~\rm{keV}$ that are incorrectly identified as background events. For this reason, the angular distribution of the left panel of Figure~\ref{fig:8keV_spectrumAngle} is truncated for angles larger than $35^\circ$ whereas we would expect the distribution to be centered on $45^\circ$. Nevertheless, the selected events reconstruct a neutron energy spectrum peaking around $7.5~\rm{keV}$ which is $8\%$ lower than the expected value, which is a small difference regarding the uncertainties.

We evaluate the bias and the resolution of the angle reconstruction in Figure~\ref{fig:8keV_thetaExp_cosSquared}. The left panel of the figure showing the evolution in energy of the ratio $\cos^2(\theta) / E_K^p$ indicates that the protons with kinetic energies larger than $4~\rm{keV}$ lead to the expected constant ratio centered on $1/E_n = 0.125~\rm{keV^{-1}}$. Below $4~\rm{keV}$ the ratio increases, showing that the angular resolution deteriorates. The right panel presents the difference between the reconstructed angle and the expected one. The distribution is centered near $0^\circ$ but its extremities deviate from this expected value. We explain this deviation by two phenomena. First, some background events could have passed the discrimination cuts, particularly at low energies, and would then be incorrectly assigned an IQF and neutron-proton collision kinematics, leading to an overestimation of the reconstructed energy. Second, the deviation from the expected value, in particular for $\theta > 45 ^\circ$, can indicate a bias in the angle reconstruction. We mentioned in Section~\ref{sec:highGain} that angles larger than $45^\circ$ are difficult to reconstruct since they are associated with a low time resolution on the quantities \texttt{$\rm{Z_{loss}}$} and \texttt{Ion duration}. We thus expect the uncertainty of the high gain 3D method to increase with the angles and we investigate in the following how this uncertainty affects the reconstructed neutron spectrum. 

Despite the deviations from the expected value in Figure~\ref{fig:8keV_thetaExp_cosSquared}, we observe that the majority of the events with kinetic energy larger than $4~\rm{keV}$ are located inside a 2D Gaussian with the $y$ component centered in $0^\circ$ and having a width of $16^\circ$. We thus consider that we measured an angular resolution of $16^\circ$ for proton recoils in the range $[4~\rm{keV}, 8~\rm{keV}]$. 

\subsection{Bi-energetic neutron field}

The situation for the 2023 campaign differs from the two previous ones. The nuclear reaction on the LiF target produces a larger proportion of neutrons compared to the gammas than the reaction on the scandium target, hence simplifying the electron-recoil discrimination procedure (\textit{c.f.} Appendix~\ref{app:eRdiscri}) and thus giving access to a lower detection threshold than in the two previous campaigns. 

\begin{figure}[t]
	\centering
	\includegraphics[width=0.8\linewidth]{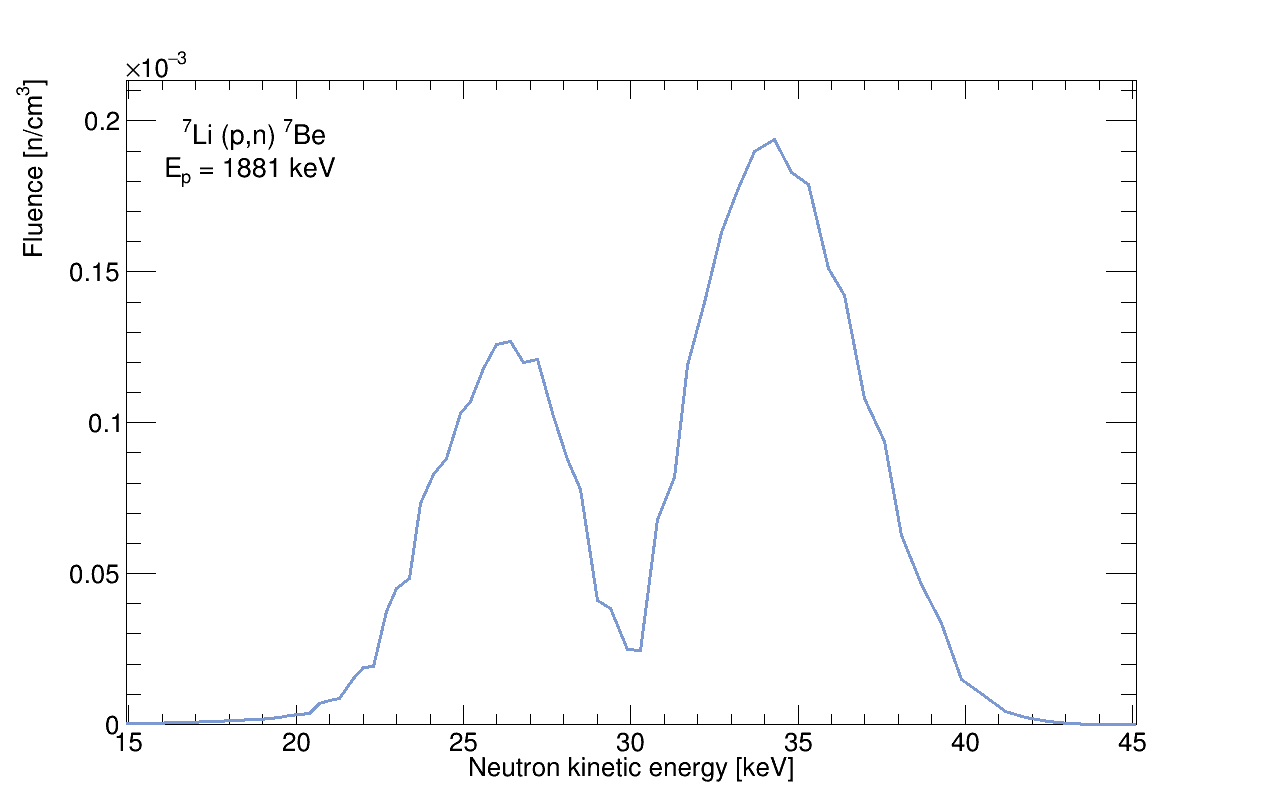}
	\caption{Simulation of the neutron fluence at the detector position as a function of the neutron energy in the conditions of the AMANDE 2023 campaign when sending a proton energy of $1881~\rm{keV}$ on the LiF target.}
	\label{fig:2023_simulatedSpectrum}
\end{figure}

The other main difference is that the neutron field is bi-energetic. We use the TARGET code \cite{TARGET} to perform a Monte Carlo simulation of the neutron fluence, as a function of the neutron energy, at the position of the detector for the experimental configuration of the 2023 AMANDE campaign. To characterize in the simulation the neutron distribution emitted around the target, the distribution is scanned from $0^\circ$ to $5^\circ$ in $1^\circ$ steps. The angular distribution is defined based on the SI and SP cards (Source Information and Source Probability). The first card is associated with the calculated angle definitions. The second card is associated with the probability density of these angles extracted from TARGET. The simulated neutron fluence at the detector position, as a function of the neutron energy, is represented in Figure~\ref{fig:2023_simulatedSpectrum}. It has two contributions: one centered on $26.0~\rm{keV}$ with a FWHM of $4.6~\rm{keV}$, and the other centered on $34.4~\rm{keV}$ with a FWHM of $5.6~\rm{keV}$. 

\begin{figure}[t]
	\centering
	\begin{minipage}{0.49\linewidth}
		\includegraphics[width=\linewidth, height=6cm]{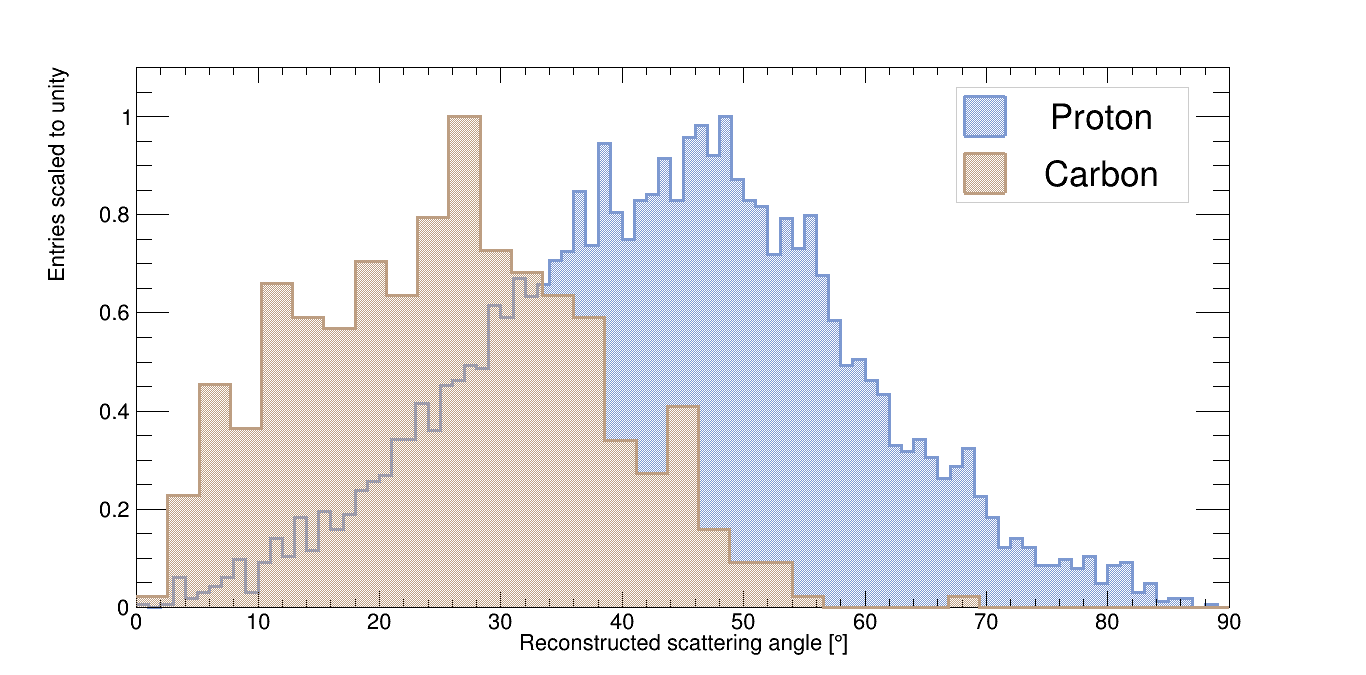}
	\end{minipage}
	\hfill
	\begin{minipage}{0.49\linewidth}
		\includegraphics[width=\linewidth, height=6cm]{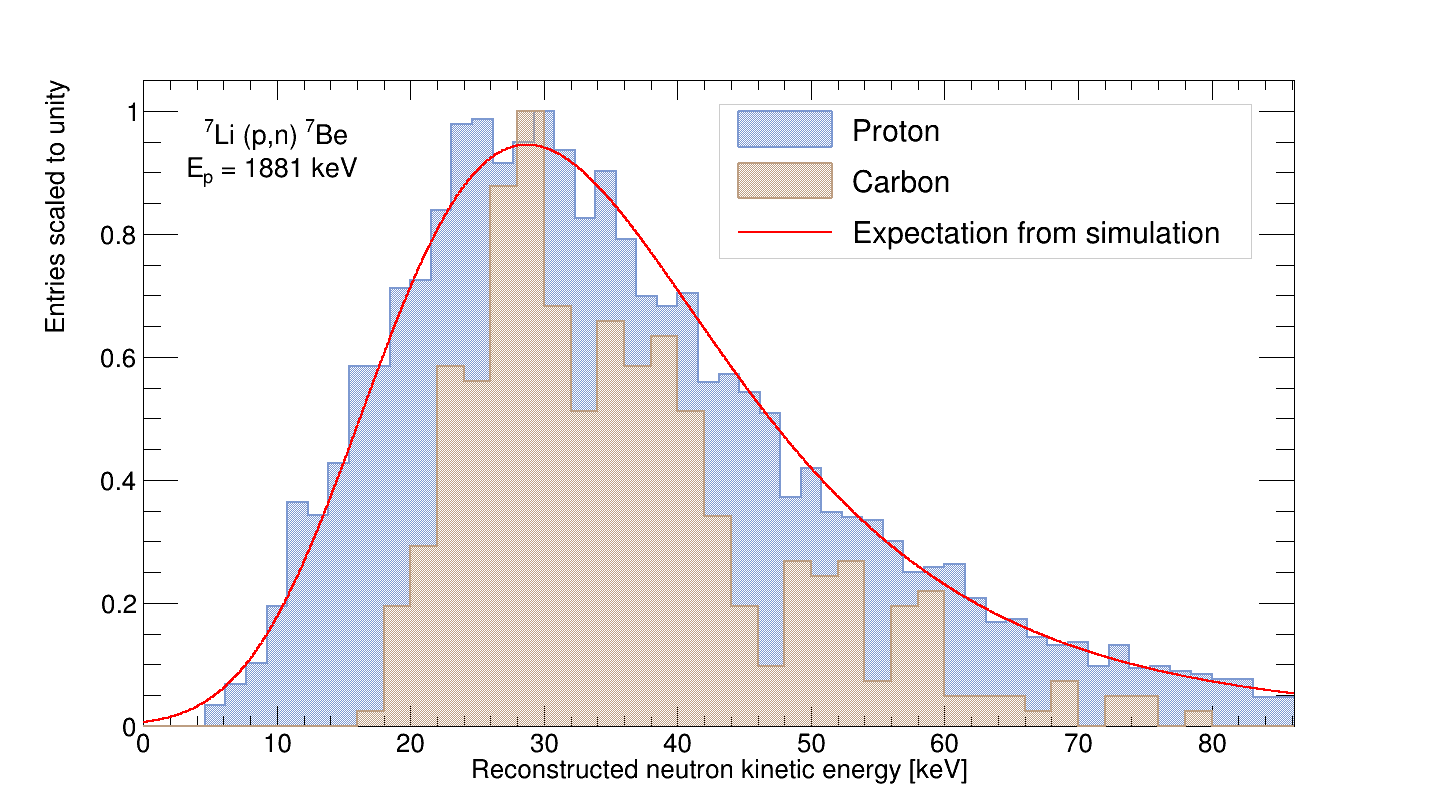}
	\end{minipage}
	\caption{Angular distribution (left) and neutron energy spectrum (right) reconstructed from the proton and carbon recoils detected in the 2023 campaign. On the right panel, the red curve is a fit of the spectrum reconstructed from the proton recoils based on the simulated spectrum of Figure~\ref{fig:2023_simulatedSpectrum}. The characteristics of the simulated spectrum are fixed in the fit, the free parameters being only associated with the uncertainties. The free parameters account for the $10\%$ uncertainty on the peaks' positions, related to the uncertainty on the energy of the proton beam of the AMANDE facility, and for an energy-dependent resolution of the MIMAC detector, modeling a possible bias of the high gain 3D method for large angles. The figures are constructed from 8531 proton recoils and 443 carbon recoils.}
	\label{fig:2023_measurements}
\end{figure}

The measurements of the scattering angles as well as the reconstructed neutron spectrum, both from proton and carbon recoils, are presented in Figure~\ref{fig:2023_measurements}. Once again, we use the $\gamma$ and $\delta$ parameters fixed from the proton recoils of the $27~\rm{keV}$ campaign. We measure an angular distribution for the proton recoils centered in $45^\circ$, as expected, while the angular distribution of the carbon recoils is truncated because of the detection threshold that eliminates carbon recoils with kinetic energies below $5~\rm{keV}$. 

The neutron spectrum in the right panel of Figure~\ref{fig:2023_measurements} requires some comments. Let us first remark that the spectra reconstructed from the proton and the carbon recoils are in agreement. One can observe that the carbon-spectrum has an overpopulation for a neutron energy near $28~\rm{keV}$ compared to the proton-spectrum. As mentioned in Appendix~\ref{app:pCdiscri}, in this campaign the neutron energy is sufficiently large to produce some fluorine recoils above the detection threshold. We have however not been able to discriminate them from the carbon recoils, and we estimate the proportion of fluorine recoils to represent about $10\%$ of the events identified as "carbon recoils". Due to the neutron-recoil kinematics Eq.~\eqref{eq:kinematics} and the IQF, a fluorine recoils incorrectly identified as a carbon recoil would be placed at an energy about $45\%$ too low in the reconstructed spectrum of Figure~\ref{fig:2023_measurements}. The misidentification of fluorine recoils as carbon recoils results then in an incorrect overpopulation of the low-energy part of the reconstructed neutron spectrum.

Let us now focus on the spectrum reconstructed from the proton recoils and compare it with the spectrum simulated with TARGET. Because of the experimental resolution, we do not expect to observe the two spectral contributions of Figure~\ref{fig:2023_simulatedSpectrum} but rather an overlap of them. Usually, we model the MIMAC resolution by a Gaussian. However, since the high gain 3D method suffers from uncertainties increasing with the angle, we decide to parametrize the width of the Gaussian function as a function of the neutron energy. We then model the resolution for the reconstructed neutron spectrum as:
\begin{equation}
	R(E_n) = e^{-\frac{E_n^2}{2\sigma(E_n)^2}} \hspace*{1.2cm} \mathrm{with}\hspace*{1.2cm} \sigma(E_n) = \sigma_0 + A\,E_n^{B}
	\label{eq:resolution}
\end{equation}

We fit the reconstructed spectrum in a restrictive way: the characteristics of the simulated spectrum of Figure~\ref{fig:2023_simulatedSpectrum} are imposed in the fit, \textit{i.e.} the amplitudes of the peaks, their positions and their widths. The fit function has five parameters: an overall normalization not playing any physical role; the three parameters of Eq.~\eqref{eq:resolution} accounting for the detector resolution; and a 10\% degree of freedom on the positions of the peaks. This $10\%$ uncertainty on the positions of the peaks, according to the TARGET simulation, corresponds to the propagation of the $1~\rm{keV}$ uncertainty on the energy of the proton beam of the AMANDE facility. 

The fit is represented as a red curve in Figure~\ref{fig:2023_simulatedSpectrum} and the best-fit results are obtained for $\sigma_0 = 8.0 \pm 0.47$, $A = 8.2\times 10^{-2} \pm 1.6\times 10^{-2}$, and $B = 1.2 \pm 0.08$. In other words, the resolution for reconstructing the neutron energy deteriorates by a factor of 2 from $E_n=0~\rm{keV}$ to $E_n=45~\rm{keV}$. This fit is an additional indication that the resolution of the high gain 3D method deteriorates when the angle is large, in particular for $\theta > 45^\circ$. However, even accounting for this bias, we are able to reconstruct a neutron spectrum consistent with the TARGET simulation from proton recoils having an angular distribution centered on $45^\circ$. 

Note that since the neutron field is bi-energetic, we cannot determine the angular resolution of the reconstruction as done in the two mono-energetic neutron campaigns. Besides the evaluation of the performance of a directional dark matter detector, the measurement of the neutron spectrum of the $^7$Li$(p,n)^7$Be reaction near threshold is of great interest as input data to design an optimized moderator for the production of a metrological reference epithermal neutron field.

\section{Discussion}\label{sec:directional}

As summarized in Table~\ref{tab:summary}, we are able to reconstruct the correct neutron energy spectrum, both from proton recoils and carbon recoils, in three different datasets, measured at three different gains, after applying three different electron-recoil discriminations. Although these combined observations are not sufficient to prove in the general case the validity of the high gain 3D method, they validate its application for proton recoils with kinetic energies in the range $[2, 40 ~\rm{keV}]$ and carbon recoils in the range $[5, 15 ~\rm{keV}]$. Moreover, they provide a quantification of the angular resolution of the MIMAC detector, in the shorter energy range tested under the mono-energetic neutrons fields, which accounts for the experimental systematic uncertainties and the one related to the angular reconstruction procedure that increases at large angles. 

There is no hard limit in the requirements for a directional detector, for reviews we refer to \cite{Mayet2016, Vahsen2021}. However, it has been shown that the most critical limitation is the energy threshold \cite{Billard2011}. Reducing the threshold is particularly challenging for directional detectors since recoils in the keV range are usually associated with sub-millimeter tracks. The $5~\rm{keV}$ carbon recoils reconstructed in this paper correspond to track lengths of $380~\rm{\mu m}$ before diffusion according to SRIM. Such a directional threshold on carbon recoils enables searching for WIMP masses down to $\mathcal{O}(10~\rm{GeV})$. The threshold of $2~\rm{keV}$ achieved for proton recoils enables searching for WIMP masses down to $\mathcal{O}(1~\rm{GeV})$ with the directional information. Note that for the experiments presented in this paper, the directional threshold is not limited by the track lengths but by the ionization threshold of the detector. For instance, the directional threshold of $2~\rm{keV}$ for proton recoils corresponds to the track lengths of $920~\rm{\mu m}$ according to SRIM, which is more than twice larger than the track length of the lowest carbon recoils that has been reconstructed. 

The directional procedure that we used in our previous paper \cite{Beaufort2021}, from the Flash-ADC, required the neutrons (or the WIMPs) to be aligned with the drift direction of the detector. The method presented in this paper does not suffer from such a geometrical restriction. To the best of the authors' knowledge, this work then achieves for the first time in a general configuration the directional measurements of proton and carbon recoils with kinetic energies in the keV range. In other words, this work demonstrates that directionality is around the corner for probing WIMPs with masses down to $1~\rm{GeV}$. The measured angular resolution is better than the usual requirement of $30^\circ$ \cite{Vahsen2021} although additional measurements in the future could provide a finer quantification of its deterioration for the angles larger than $45^\circ$.

 Most of the directional detectors\footnote{The only exception with published results is the NEWSdm experiment \cite{Asada2022} which is based on nuclear emulsions in a solid target made of AgBr crystals dispersed in a polymer.} are gaseous TPCs operating at low pressure (for instance $30~\rm{mbar}$ for MIMAC) to increase the track lengths of the nuclear recoils, thus reaching low densities compared to the ones of non-directional detectors, hence low exposures. However, this argument can be placed in perspective by at least two elements that we propose as an interesting challenge for coming years. Recently, the NEWS-G collaboration has presented non-directional preliminary results on the spin-dependent WIMP-proton cross-section \cite{Paco2022} which would correspond to the world-leading constraints on WIMP masses in the range $0.2-2~\rm{GeV}$. NEWS-G also exploits a gaseous TPC and reads an ionization signal. The reported effective exposure is $0.12~\rm{kg\cdot day}$ corresponding to 10 days of measurements in $135~\rm{mbar}$ of pure methane in a volume of $1.4~\rm{m^3}$. These preliminary results indicate that the relatively low exposure of directional detectors, which can be of the order of the one used by NEWS-G, could contribute exploring new regions of the WIMP parameter space. Finally, we mention that directional detectors require lower exposures than non-directional detectors to explore the neutrino fog due to the different angular distributions of the WIMP and the neutrino induced recoils \cite{OHare2015}.

The second element of interest concerning the discovery reach of directional experiments is the use of light targets (H, He, C, F, etc.) in directional detectors in order to increase the track lengths of the nuclear recoils. While the spin-independent (SI) WIMP-nucleus cross-section scales with the target atomic number squared, hence favoring heavy targets such as xenon for instance, this is not the case for the spin-dependent (SD) cross-section \cite{Cerdeno2011}. In the minimal supersymmetric standard model (MSSM), the SD contribution dominates over the SI one up to atomic numbers $A\simeq 20$ \cite{Jungman1995}, reaching three orders of magnitude difference for the H target. Note also that the kinematics of the WIMP-nucleus scattering favours target nuclei of masses similar to the WIMP mass. Finally, the event rate exponentially increases at low energy so a key point in WIMP detection consists in reducing the energy threshold. For detectors relying on the ionization signal, either for the energy measurement or for the electron-recoil discrimination, using a light target reduces the impact of ionization quenching. 

We conclude this discussion by commenting on two phenomena playing an important role in the directional detection of sub-millimeter tracks. It has long been thought that diffusion was an issue for gaseous directional detectors \cite{Sciolla2009, Vahsen2020}. We claim the opposite. For WIMP masses below $30~\rm{GeV}$, the WIMP-induced recoils would have track lengths smaller than one millimeter. Without diffusion, such tracks would be detected on a few pixels so that the angular resolution would be significantly limited by the dimensions of the pixel. In other words, to measure the directional information, the track length must be large compared to the dimension of the readout element. By enlarging the primary electrons cloud the diffusion increases the number of fired pixels and the signal-over-threshold time. We stress that the diffusion enlarges the cloud in a Gaussian way, hence most charges remain in the center of the track while the tail improves the resolution. For sub-millimeter tracks, a balance must be found between too much diffusion, which at some point deteriorates the directional information, and too little diffusion. In this work, we measure the direction of carbon recoils down to $5~\rm{keV}$ (corresponding to track lengths down to $380~\rm{\mu m}$ before diffusion according to SRIM) and we are limited to access lower energies by the detection threshold, not by diffusion. We thus claim that diffusion coefficients of the order of $350~\rm{\mu m / \sqrt{cm}}$ and an active volume having a maximal drift of $25~\rm{cm}$ do not represent a limitation, and we rather consider that in such configuration the diffusion represents a key help for such directional measurements. 

The second phenomenon influencing the directional detection of sub-millimeter tracks corresponds to the high gain effects detailed in Section~\ref{sec:highGain}. In a fashion similar to the discussion about the diffusion, we consider that the improved sensitivity offered by the high gain effects more than compensates the disadvantages due to the track distortions. The baseline produced by the ionic signal, even if it distorts the tracks, provides two crucial contributions: it reduces the detection threshold and it acts as a zoom through the track by time-spreading. Once handled, either by the analytical deconvolution of the Flash-ADC signal or by the high gain 3D method, it becomes possible to exploit this additional sensitivity without suffering from the distortions. The demonstrated angular resolution of $16^\circ$ for proton and carbon recoils with kinetic energies down to a few keVs indicates that the directional information  is preserved by the high gain effects. The description and the simulation of the high gain effects observed in MIMAC, mainly developed in \cite{Beaufort2021, ThesisCyprien}, could possibly be extended to other directional detectors not relying on Micromegas.

\section{Conclusion}

Directional detection is the appropriate strategy for WIMP discovery and directional experiments can explore new regions of the WIMP parameter space, as mentioned in the previous section, even accounting for their relatively low exposures compared to the ones of non-directional detectors. By measuring a better than $16^\circ$ angular resolution on keV proton and carbon recoils with MIMAC, the current work opens the window for directional WIMP searches with masses down to $\mathcal{O}(1~\rm{GeV})$. The directional method proposed in this paper depends however on experimentally fixed parameters and must be tested by exposing the detector to neutron fields, for each gas mixture, in order to determine its validity over a given energy range and to quantify its angular resolution. 

The directional performance presented in this work embeds all the experimental uncertainties such as the ones related to the calibration, to the Ionization Quenching Factor, to the discrimination electron-recoil, to the discrimination proton-carbon, and to the angle reconstruction method. This work can thus be viewed as an overall validation of the MIMAC strategy of detection of keV nuclear recoils. Such a validation was a mandatory step before performing runs for WIMP searches. For this purpose, a bi-chamber module of the MIMAC prototype with an active volume of $6~\rm{L}$ is operating at the Underground Laboratory of Modane since March 2023. 

The main limitation of the present work comes from the ionization threshold of the MIMAC detector. We have developed a new detector based on the resistive Micromegas technology \cite{Alexopoulos2011, Galan2013} to reach a lower threshold. We have recently demonstrated a $28~\rm{eV}$ ionization threshold \cite{ThesisCyprien} while reconstructing the spectrum and the tracks of $150~\rm{eV}$ electrons sent in the detector by Comimac. We are currently testing such a resistive detector having an anode of $35\times35~\rm{cm^2}$, which would improve by one order of magnitude the volume of the MIMAC detector. By extending to resistive Micromegas the method presented in this paper, we expect to significantly improve in the near future the directional performance of the MIMAC detector.

\newpage
\appendix
\section*{Appendices}
\addcontentsline{toc}{section}{Appendices}
\renewcommand{\thesubsection}{\Alph{subsection}}
\renewcommand{\theequation}{\Alph{subsection}.\arabic{equation}}

\subsection{Calibration}\label{app:calib}

For reproducibility, the gas mixture is placed into a buffer volume used to fill the detector once a day during the entire measurement campaign. The calibration is performed before the campaign and checked afterwards to control the gas stability.

The calibration is performed with the table-top accelerator Comimac coupled to a MIMAC chamber to send electrons of controlled kinetic energy in the detector, more details can be found in \cite{Muraz2016, NEWS-G2022}. The ionization energy released by the electrons, which stop in the MIMAC chamber, is proportional to the amplitude of the signal measured by a charge-sensitive preamplifier coupled to the Flash-ADC. For each energy, we fit the histogram of the amplitude distribution by a Gaussian. We apply a single cut to the data: since Comimac sends the electrons at the center of the detector, we reject the events located out of the central region which are due to the cosmic background. 

\begin{figure}[t]
	\centering
	\includegraphics[width=0.8\linewidth]{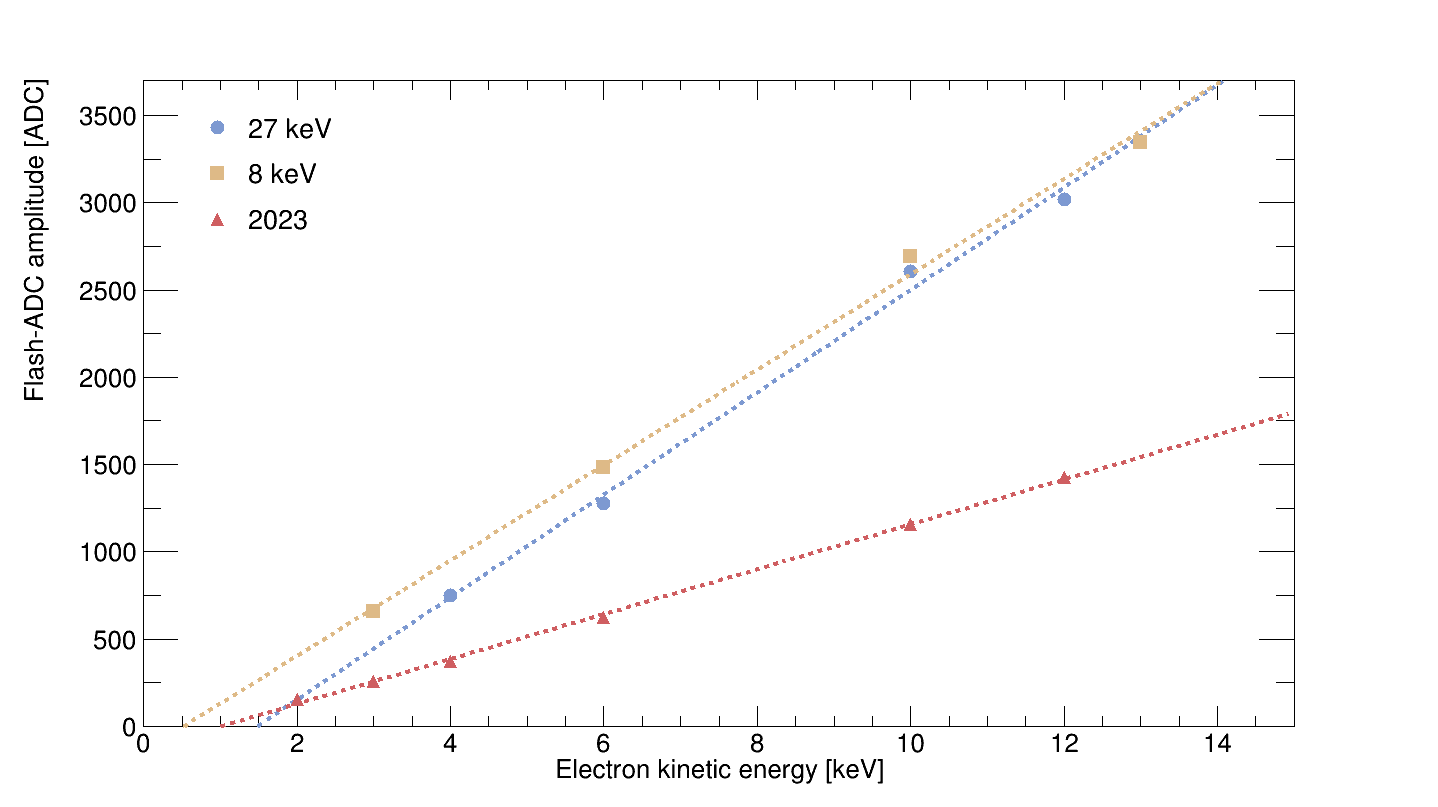}
	\caption{Calibrations for the measurement campaigns on AMANDE.}
	\label{fig:calibration}.
\end{figure}

In Figure~\ref{fig:calibration} we present the calibrations obtained for the three measurement campaigns with neutron fields. Such calibrations are consistent with a linear detector response in the $[2,13]~\rm{keV}$ range. Each calibration presents an offset related to the ambient electromagnetic noise and to the electronic chain, which then depends on the experimental conditions. The offset contributes significantly to the detection threshold and limits the ability to access to the sub-keV region.

\subsection{Ionization Quenching Factor}\label{app:IQF}

The amount of energy released by a nuclear recoil ionizing the atoms of the active volume of detection appears "quenched" compared to an electron of the same kinetic energy. To describe this behavior, we introduce the Ionization Quenching Factor (IQF) defined as:
\begin{equation}
	\rm{IQF}(E_K) = \frac{E_I^{recoil}}{E_I^{elec.}}\bigg|_{E_K}~,
\end{equation}
that is the ratio of the ionization energy measured for a recoil over the one measured for an electron of the same kinetic energy $E_K$. 

The IQF is a crucial quantity to reconstruct the kinetic energy of a nuclear recoil when measuring its ionization. It depends on the recoil mass and on its kinetic energy, but also on the gas mixture, the pressure, and a calibration to evaluate the energy of the electrons. 
 
To the best of the authors' knowledge, there is no appropriate theoretical modelling of the IQF in gas mixtures at the moment. The \texttt{SRIM} simulation toolkit \cite{SRIM} relies on the Lindhard theory \cite{Lindhard1963} which defines a quenching factor as the fraction of the kinetic energy ultimately given to the electrons of the medium from inelastic collisions resulting in the ejection of electrons or in electronic excitations. In this sense, the Lindhard quenching also embeds the scintillation and corresponds to an upper limit for the IQF \cite{Sorensen2015, Sarkis2020}. 

The measurement of the IQF is thus a mandatory step for neutron spectroscopy and WIMP searches with an ionization detector such as MIMAC. We use Comimac to send electrons and ions of the same kinetic energy to measure the IQF. Note that ions and nuclear recoils have the same IQF \cite{SRIM} due to the continuously changing state of charge during the ionization process. We parametrize the IQF in our gas mixture i-C$_4$H$_{10}$ + 50\% CHF$_3$ at $30~\rm{mbar}$ such as:
\begin{equation}
	\rm{IQF} (E_K) = N\,E_K^\alpha / (\beta + E_K^\alpha)~.
	\label{eq:IQFparam}
\end{equation}

For the proton IQF, we have performed additional measurements since the publication of \cite{Beaufort2021} leading to an update of the parameters to $N=1$, $\alpha = 0.38 \pm 0.04$, and $\beta = 1.43 \pm 0.11$. The measurement of the carbon IQF with Comimac is however challenging because the carbon ions sent by Comimac make a deposit in the interface between Comimac and the MIMAC chamber and obstruct it. In other words, the measurement deteriorates the setup. For this reason, we have performed a carbon IQF measurement for a single energy, obtaining an IQF of $0.40$ for $10~\rm{keV}$ carbon ions. Since we do not have a complete carbon IQF measurement, we proceed by an approximation: we normalize the Lindhard quenching simulated by \texttt{SRIM} such that it equals $0.40$ at $10~\rm{keV}$. In this configuration, we obtain the following parameters for the carbon IQF: $N=0.8$, $\alpha = 0.41 \pm 0.02$, and $\beta = 2.43 \pm 0.10$.

\subsection{Electron-recoil discrimination}\label{app:eRdiscri}

The majority of the events measured during the AMANDE campaigns are due to the background. For instance, the nuclear reaction on the $^{45}$Sc target produces photons with a fluence 20 times larger than the one for the neutron production \cite{Maire2016}. The gamma and the muon represent the dominant background sources in our measurements and they result in electron-like signals in the detector. For measurements of a couple of hours and in the energy range considered, the neutron background is negligible with respect to the fluence of the neutrons sent by AMANDE.

At high gain and large gap, the influence of the ionic signal described in Section~\ref{sec:highGain} makes the electron-recoil discrimination more complex. For each campaign, we thus performed specific background measurements in varying the energy of the protons beam to be either below the neutron production threshold (for the 2023 campaign), either to be out of the resonance for the neutron production. Such "background only" datasets are used for the electron-recoil discrimination by comparing with the standard measurements that contain both signal and background. 

Recoil tracks form a single and dense cluster observable both on the anode and on the Flash-ADC. While electron signals can also present the same features, the opposite is not true. We thus apply a set of \textit{minimal cuts} to reject events that are incompatible with recoil tracks. These minimal cuts are specific to MIMAC but we list them here for completeness. They can be grouped into four categories:
\begin{itemize}
    \item The derivative of the Flash-ADC has a single peak that contains more than 96\% of the total energy deposition. The baseline must have fluctuations smaller than 25 ADC and its first value must be lower than 400 ADC.
    \item The track must be entirely contained into the detection volume and its must have at least a strip fired in two different timeslices in order to provide a direction.
    \item To reject failure of the Flash-ADC deconvolution, the duration of the primary electrons cloud must be at least larger than half of the ion duration. As explained in Section~\ref{sec:highGain}, the high gain 3D method requires a large enough pixel density tested by a ratio between the number of activated pixels projected in 2D divided by $\Delta X \cdot \Delta Y$ that must exceed 0.025.
    \item For the mono-energetic neutron fields, the nuclear recoils populate a branch on a 2D histogram of the track length as a function of the ionization energy. We add a cut to remove events that have obviously too long track lengths with respect to the energy.
\end{itemize}

When applying the minimal cuts to the 2023 campaign with the bi-energetic neutron field, based on the $^7$Li$(p,n)^7$Be reaction, 7.26\% of the events are kept in the "signal + background" dataset whereas only 0.39\% events are kept in the "background only" dataset. By comparing the two datasets, we expect the electron background to represent about 5\% of the kept events in the analysis after applying the minimal cuts. Such an electron-recoil discrimination is sufficient for the analyses presented in Section~\ref{sec:measurements}.

However, for the mono-energetic $27~\rm{keV}$ and $8~\rm{keV}$ neutron campaigns, which are based on the $^{45}$Sc$(p,n)^{45}$Ti reaction, the minimal cuts make only a difference of 0.2\% between the "background only" and the "background + signal" data. We thus go one step further by training Boosted Decision Trees (BDT) to identify the signal based on 12 discriminating observables. Note that the discriminating observables must be independent of the quantities \texttt{$\rm{Z_{loss}}$} and \texttt{Ion duration} used to reconstruct the angle as detailed in Section~\ref{sec:highGain}.

The procedure for the BDT applied to the $27~\rm{keV}$ dataset is detailed in \cite{ThesisCyprien}. The price to pay when using the BDT is that part of the events (in this case 25\%) are used to train and test the machine learning algorithm and are excluded for the rest of the analysis in order to reduce bias. For each event, the BDT provides a probability for the event to be a recoil. To reduce the electron events in the final data until a negligible quantity, we place a stringent cut on the probability returned by the BDT which removes at the same time about $50\%$ of the events. In other words, while the BDT is a powerful tool to reject the background, it significantly reduces at the same time the number of kept recoil events. 

Finally, we ran into difficulties for the $8~\rm{keV}$ campaign since the lower the energy, the more difficult to separate an electron from a recoil. Our first BDT tests introduced a bias in energy so we decided to stick with manual discrimination without BDT in our previous paper \cite{Beaufort2021}. We have eventually been able to use a BDT as follows: we train and test the BDT on the events selected for the previous paper, and then we apply it to the totality of the events passing the minimal cuts.

\begin{figure}[t]
	\centering
	\includegraphics[width=0.8\linewidth]{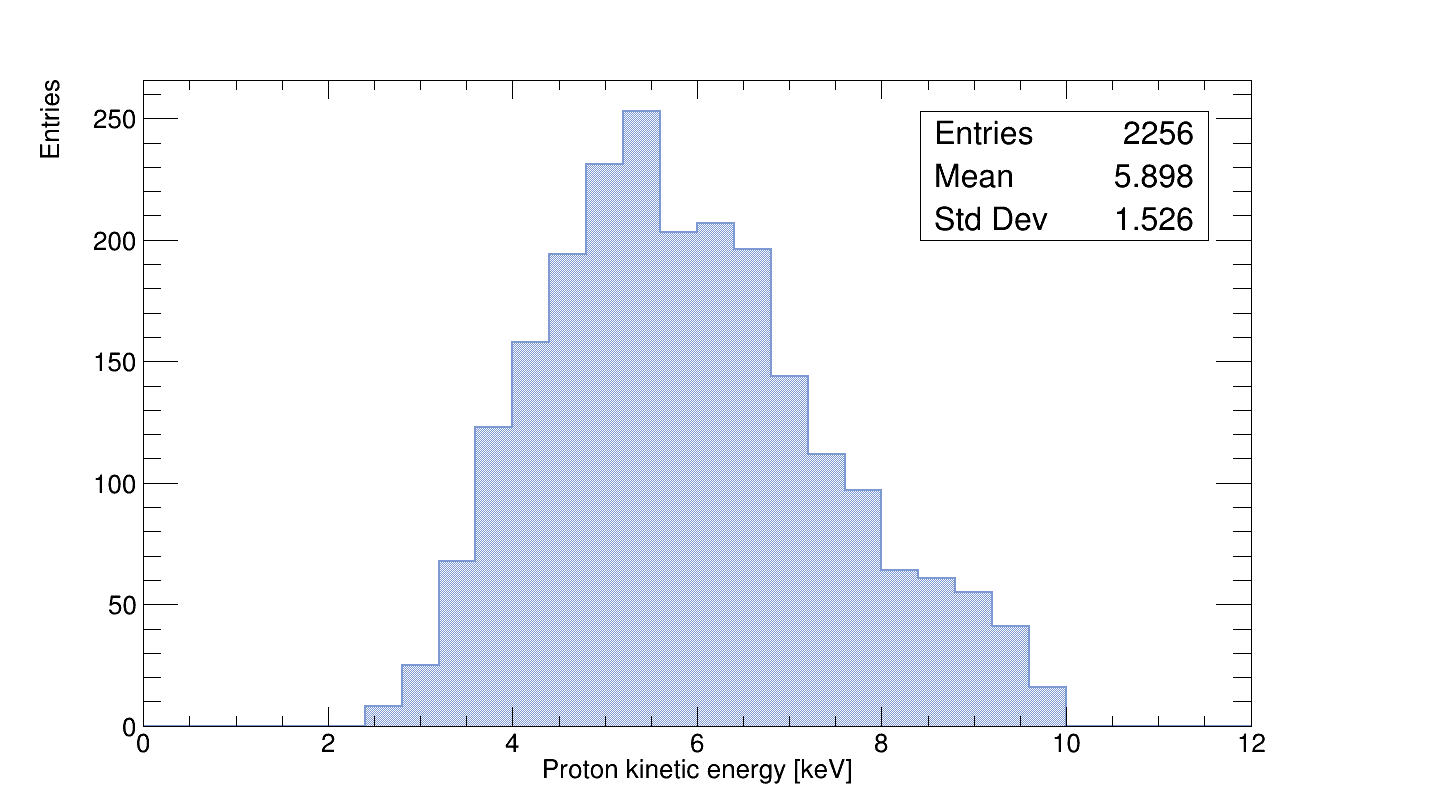}
	\caption{The spectrum of the kinetic energy of the proton recoils, when accounting for the IQF, in the AMANDE $8~\rm{keV}$ campaign after applying the BDT.}
	\label{fig:app8keVspectrum}
\end{figure}

Figure~\ref{fig:app8keVspectrum} presents the energy spectrum of the corresponding proton recoils selected by the BDT for the $8~\rm{keV}$ campaign. The kinetic energy is reconstructed by the calibration and the IQF. The events exceeding $8~\rm{keV}$ are background events not detected by the BDT. Only a few events are detected below $4~\rm{keV}$ which is a bias from the BDT that is rarely able to separate the electrons from the recoils with a high enough probability at such low energies, so the doubted events are rejected. This low-energy effect will have repercussions on the angular distribution of the proton recoils that will be partially truncated for angles larger than $45^\circ$. While these identified biases with the BDT limit the detection, we see in Section~\ref{sec:measurements} that the angular resolution of the proton recoils remains at an acceptable value of $16^\circ$.

\subsection{Proton-carbon discrimination}\label{app:pCdiscri}

Once we have identified the nuclear recoils in the data, the next step consists of separating the carbon recoils from the proton recoils. Our gas mixture also contains fluorine atoms, so fluorine recoils are expected. However, due to the kinematics and the IQF, they will not exceed the detection threshold for the two mono-energetic datasets, $27~\rm{keV}$ and $8~\rm{keV}$. For the 2023 campaign, which reaches larger energies, we expect to detect a few tens of fluorine recoils,  about 10 times less than carbon recoils. Due to these low statistics, we have not been able to discriminate the fluorine from the carbon recoils. In other words, some fluorine recoils populate what we call the "carbon recoils" although with a minor influence. 

To discriminate the carbon and the proton recoils, we proceed similarly as for the electron-recoil discrimination: we first apply a manual cut to reject events that are necessarily proton recoils, then we use a BDT. Since the carbon recoils are denser than the proton recoils we define a figure of merit (FoM) to select the dense tracks:
\begin{equation}
	\rm{FoM} ~=~ \Delta X \times \Delta Y \times \Delta T \times (\rm{NbHoles} X +1) \times (\rm{NbHoles} Y +1) \times (\rm{NbHoles} T +1) ~,
\end{equation}
in which the $+1$ avoids multiplying by zero. The lower the FoM, the denser the track. Due to diffusion, a proton recoil produced near the MIMAC grid can be as dense as a carbon track that has drifted over the entire active volume. A dense track is thus not a sufficient observable to identify the carbon, but a non-dense recoil track is necessarily associated to a proton track. We can also use the kinematics Eq.~\eqref{eq:kinematics} to determine the maximal kinetic energy of the carbon recoils, the so-called \textit{endpoint}, which is lower than the one for the proton recoils. Having this information in mind, we define the maximal allowed FoM equal to its value at the carbon endpoint energy. 

At this stage, we train and test a BDT that now uses the FoM-selected events as the "signal + background". We apply this BDT on the events passing the minimal cuts. In other words, the carbon-proton BDT and the electron-recoil BDT are trained differently but they are applied on the same dataset. 

In the 2023 campaign, we identified $479$ carbon-like events corresponding to $0.1\%$ of the whole events and $6\%$ of the recoil events. In the $27~\rm{keV}$ campaign, we identify as many carbon recoils than proton recoils, probably due to the more difficult electron-recoil discrimination, compared to the 2023 campaign, which rejects the majority of proton recoils whereas carbon recoils are denser so easier to discriminate from electrons. In the $8~\rm{keV}$, the carbon endpoint energy is comparable to the detection threshold so we do not expect to observe carbon recoils in the data.

\bibliographystyle{JHEP}
\bibliography{references}

\end{document}